\RequirePackage{fix-cm}
\documentclass[twocolumn,epjc3]{svjour3}  
\usepackage{newtxtext,newtxmath} % If using this, comment out mathptmx and upgreek down below
\RequirePackage{graphicx}
\RequirePackage{subfigure}
\RequirePackage{rotating}
\RequirePackage{pdflscape}
\RequirePackage{color}
\usepackage{mathtools}
\usepackage{extarrows}
\usepackage{bm}
\RequirePackage{latexsym}
\RequirePackage{xspace}
\RequirePackage[numbers,sort&compress]{natbib}
\RequirePackage[colorlinks,citecolor=blue,urlcolor=blue,linkcolor=blue]{hyperref}
\usepackage{amsmath,amsfonts,bm}
\usepackage{wasysym}
\usepackage{color}
\usepackage{lineno}
\usepackage[usenames,dvipsnames,svgnames,table]{xcolor}	%colored text
\let\oldequation\equation
\let\oldendequation\endequation

\renewenvironment{equation}
  {\linenomathNonumbers\oldequation}
  {\oldendequation\endlinenomath}

\let\oldalign\align
\let\oldendalign\endalign

\renewenvironment{align}
  {\linenomathNonumbers\oldalign}
  {\oldendalign\endlinenomath}

\usepackage{acronym}
\usepackage{xcolor}
\usepackage{braket}

\journalname{Eur. Phys. J. C}
\begin{document}

\title{The CROSS experiment: detector construction, background projection, and sensitivity to $^{100}$Mo $0\nu2\beta$ decay}

\author{
D.~Auguste\thanksref{IJCLab}\and
%I.~Bandac\thanksref{LSC}\and
A.S.~Barabash\thanksref{KCTEP}\and
G.~Benato\thanksref{GSSI,LNGS}\and
V.~Berest\thanksref{CEA-IRFU,e1}\and
L.~Berg\'e\thanksref{IJCLab}\and
M.~Buchynska\thanksref{IJCLab,e3}\and
J.M.~Calvo-Mozota\thanksref{LSC}\and
J.~Cao\thanksref{IJCLab,e2}\and
P.~Carniti\thanksref{INFN-Milano}\and
M.~Chapellier\thanksref{IJCLab}\and
D.~Cintas\thanksref{IJCLab,CEA-IRFU,e4}\and
I.~Cojocari\thanksref{IJCLab}\and
I.~Dafinei\thanksref{INFN-Roma}\and
F.A.~Danevich\thanksref{KINR,IEAP}\and
M.~De~Deo\thanksref{LNGS}\and
A.~Drobizhev\thanksref{LBNL}\and
L.~Dumoulin\thanksref{IJCLab}\and
F.~Ferri\thanksref{CEA-IRFU}\and
A.~Giuliani\thanksref{IJCLab,e5}\and
C.~Gotti\thanksref{INFN-Milano}\and
Ph.~Gras\thanksref{CEA-IRFU}\and
A.~Ianni\thanksref{LNGS}\and
V.V.~Kobychev\thanksref{KINR}\and
Yu.~G.~Kolomensky\thanksref{UCB,LBNL}\and
S.I.~Konovalov\thanksref{KCTEP}\and
P.~Loaiza\thanksref{IJCLab}\and
P.~de~Marcillac\thanksref{IJCLab}\and
S.~Marnieros\thanksref{IJCLab}\and
C.A.~Marrache-Kikuchi\thanksref{IJCLab}\and
M.~Martinez\thanksref{CAFAE} \and
C.~Nones\thanksref{CEA-IRFU}\and
E.~Olivieri\thanksref{IJCLab}\and
A.~Ortiz~de~Sol\'orzano\thanksref{CAFAE}\and
M.~Pageot\thanksref{CEA-IRFU}\and
Y.~Peinaud\thanksref{IJCLab}\and
G.~Pessina\thanksref{INFN-Milano}\and
D.V.~Poda\thanksref{IJCLab}\and
Ph.~Rosier\thanksref{IJCLab}\and
B.~Schmidt\thanksref{CEA-IRFU}\and
R.~Serino\thanksref{IJCLab}\and
V.I.~Tretyak\thanksref{KINR,IEAP}\and 
V.I.~Umatov\thanksref{KCTEP}\and
M.~Velazquez\thanksref{SIMAP}\and
M.~Zarytskyy\thanksref{KINR}\and
A.~Zolotarova\thanksref{CEA-IRFU}
}

\thankstext{e3}{e-mail: mariia.buchynska@ijclab.in2p3.fr}
\thankstext{e4}{e-mail: david.cintasgonzalez@cea.fr}
\thankstext{e5}{e-mail: andrea.giuliani@ijclab.in2p3.fr}
\thankstext{e1}{Now at: University of California, Berkeley, USA}
\thankstext{e2}{Also at: Fudan University, Shanghai, China}
%\thankstext{e2}{Now at: IRFU, CEA, Université Paris-Saclay, Saclay, France}

\institute{Université Paris-Saclay, CNRS/IN2P3, IJCLab, 91405 Orsay, France \label{IJCLab} 
\and
%Laboratorio Subterr\'aneo de Canfranc, 22880 Canfranc-Estaci\'on, Spain \label{LSC}
%\and
National Research Center Kurchatov Institute, Kurchatov Complex of Theoretical and Experimental Physics, 117218 Moscow, Russia \label{KCTEP} 
\and
Gran Sasso Science Institute,  I-67100 L'Aquila, Italy \label{GSSI} 
\and
INFN, Laboratori Nazionali del Gran Sasso, I-67100 Assergi (AQ), Italy \label{LNGS}
\and
IRFU, CEA, Universit\'{e} Paris-Saclay, F-91191 Gif-sur-Yvette, France  \label{CEA-IRFU} 
\and
Laboratorio Subterr\'aneo de Canfranc, 22880 Canfranc-Estaci\'on, Spain \label{LSC}
\and
%Escuela Superior de Ingenier\'ia y Tecnolog\'ia, Universidad Internacional de La Rioja, 26006 Logro\~no, Spain \label{ESIT}
%\and
INFN, Sezione di Milano-Bicocca, I-20126 Milano, Italy \label{INFN-Milano} 
\and 
INFN, Sezione di Roma, I-00185, Rome, Italy \label{INFN-Roma}
\and 
Institute for Nuclear Research of NASU, 03028 Kyiv, Ukraine \label{KINR}
\and
Institute of Experimental and Applied Physics, CTU Prague, 11000 Prague, Czech Republic \label{IEAP}
\and
Lawrence Berkeley National Laboratory, Berkeley, CA 94720, USA
\label{LBNL}
\and
University of California, Berkeley, Berkeley, CA 94720, USA\label{UCB}
\and
Centro de Astropart\'iculas y F\'isica de Altas Energ\'ias, Universidad de Zaragoza, Zaragoza 50009, Spain \label{CAFAE}
\and
Univ. Grenoble Alpes, CNRS, Grenoble INP (Institute of Engineering Univ. Grenoble Alpes), SIMAP, 38000 Grenoble, France \label{SIMAP}
}

%\titlerunning{Short form of title}        % if too long for running head

%\thankstext{t1}{Grants or other notes
%about the article that should go on the front page should be
%placed here. General acknowledgments should be placed at the end of the article.
%\thankstext{e1}{e-mail: fauthor@example.com}

%\authorrunning{Short form of author list} % if too long for running head

\date{Received: date / Accepted: date}
% The correct dates will be entered by the editor

\maketitle

\begin{abstract}
The CROSS experiment to search for neutrinoless double-beta ($0\nu2\beta$) decay in $^{100}$Mo with the help of an array of scintillating cryogenic calorimeters, containing 4.9~kg of $^{100}$Mo, has been ongoing in a low-background setup at the Canfranc underground laboratory (Spain) since mid-November 2025. In this paper, we present the construction of the CROSS detector and the description of Geant4-based Monte Carlo simulations of expected background in the region of interest. The simulations predict the background index in a 100-keV-wide interval centered at the $Q$-value of $^{100}$Mo (3034 keV) on the level of 3.2(5) $\times$ 10$^{-3}$ cnts/keV/kg/yr. Taking into account an 18\% deadtime induced by the muon veto cut and a typical 90\% duty cycle of the facility, such background level would allow to reach the world-leading sensitivity to $^{100}$Mo $0\nu2\beta$ decay (lim $T_{1/2} \sim 4 \times 10^{24}$ yr) in 1 year of data taking. Considering conservatively a factor 3 (10) worse background index due to unpredictable radioactive contamination of construction materials and/or detector performance, a 2-yr-long operation of the CROSS array would be still compatible with the best (competitive) sensitivity to $0\nu2\beta$ decay in $^{100}$Mo.
\end{abstract}

%%=======================================================
\section{Introduction}
\label{sec:intro}

Double beta decay with neutrino emission ($2\nu2\beta$), being a second-order process in weak interactions, is the rarest nuclear disintegration ever detected \cite{Pritychenko:2025,Saakyan:2013}. During about 80 years of $2\nu2\beta$ searches, it was measured for only a dozen nuclei (1/6 of candidates \cite{Tretyak:2002}) with half-lives in the range of 10$^{18}$--10$^{24}$ yr \cite{Pritychenko:2025}. Despite a ``pure'' nuclear physics nature of this process, its investigation, particularly spectral shape, allows to probe physics beyond the Standard Model (BSM) of particle physics, particularly searching for Lorentz violation, existence of right-handed weak interactions, neutrino self-interactions, sterile / bosonic neutrinos, and Majorons \cite{Augier:2024bsm,Ghinescu:2022,Bolton:2021,Deppisch:2020,Deppisch:2020neutrino,Barabash:2007bosonic}. Also, a comparison of the measured half-life value with the theoretically predicted one is an important benchmark of the nuclear models used, taking into account its application to calculations of nuclear matrix elements for another, hypothetical, neutrinoless double-beta ($0\nu2\beta$) decay. The $0\nu2\beta$ decay is a unique probe of BSM physics, since its occurrence requires the violation of the total lepton number by two units and the existence of Majorana neutrinos with finite mass \cite{Giuliani:2026,RPP:2025,Agostini:2023,GomezCadenas:2023,Bossio:2023}. Thus, most double-beta decay search experiments are primarily focused on $0\nu2\beta$ decay searches, but no signal has been observed so far. The most sensitive experiments set lower half-life limits of 10$^{24}$--10$^{26}$ yr \cite{Giuliani:2026,RPP:2025}. From the variety of candidates, only a dozen nuclei are experimentally relevant thanks to a high transition energy ($Q_{2\beta}$ $\sim$ 2--4 MeV), a relatively high isotopic abundance ($\sim$ 10\%--30\%), and the availability of a detector technology scalable to large mass \cite{Giuliani:2026,Agostini:2023,GomezCadenas:2023}.

The isotope $^{100}$Mo ($Q_{2\beta}$ = 3034 keV) is one of a few nuclei with a long history of $2\beta$ decay search experiments.  
NEMO-3 \cite{Arnold:2015} and its predecessors, NEMO-2 \cite{Arnold:2000limits} and NEMO-1 \cite{Dassie:1991NEMO}, exploiting a tracko-calorimetry technology in the Modane undeground laboratory (LSM) in France, were long-standing leading experiments of 1990s--2020s to search for $0\nu2\beta$ decay in $^{100}$Mo. The NEMO-3 set the half-life limit relative to $^{100}$Mo $0\nu2\beta$ decay of $T_{1/2} > 1.1 \times 10^{24}$ yr. 
The LUMINEU experiment \cite{Armengaud:2017, Poda:2017a} carried out in 2017 in LSM was a small-scale demonstrator of a scintillating bolometer technology based on lithium molybdate (Li$_{2}$MoO$_4$, LMO) crystals produced from molybdenum enriched in $^{100}$Mo. Despite a small exposure of $^{100}$Mo compared to NEMO-3 (0.06 kg $\times$ yr vs. 34.3 kg $\times$ yr), LUMINEU was able to precisely measure the $^{100}$Mo $2\nu2\beta$ decay half-life, while reaching an order of magnitude lower sensitivity to $0\nu2\beta$ decay. 
CUPID-Mo~\cite{Armengaud:2020a,Augier:2022} was a LUMINEU follow-up operating a 5 times larger detector array in the same EDELWEISS low-background cryogenic facility at the LSM in 2019--2020. Thanks to increased exposure (1.47 kg $\times$ yr), high detector performance, and very low background, CUPID-Mo was able to set a new limit on the $^{100}$Mo $0\nu2\beta$ decay half-life: $T_{1/2} > 1.8 \times 10^{24}$ yr \cite{Augier:2022}. This result has been recently improved to $T_{1/2} > 2.9 \times 10^{24}$ yr \cite{Agrawal:2025amoreI} by AMoRE-I experiment (3.89 kg $\times$ yr)\footnote{Its predecessor, AMoRE-Pilot, operated 6 calcium molybdate crystals \cite{Alenkov:2019jis} and achieved an order of magnitude lower sensitivity with 6 times lower $^{100}$Mo exposure (0.68 kg $\times$ yr) \cite{Agrawal:2024}.} exploiting $^{100}$Mo-enriched scintillating low-temperature detectors based on 13 calcium molybdate and 5 LMO crystals instrumented with metallic magnetic calorimeters. In 
addition, the LMO crystal scintillator is chosen as a detector material for two large-scale cryogenic experiments, CUPID \cite{CUPID_baseline:2025} and AMoRE-II \cite{Agrawal:2025projected}, to search for $0\nu2\beta$ decay in $^{100}$Mo; data collection of the first stage of both experiments is expected to start in early 2030s.

CROSS (Cryogenic Rare-event Observatory with Surface Sensitivity) \cite{Bandac:2020} is an ERC-funded project aiming to develop metal-coated thermal detectors capable of particle identification of near surface interactions and to use them in a high-sensitivity search for $^{100}$Mo $0\nu2\beta$ decay at the Canfranc underground laboratory (LSC) in Spain. Following the purification and crystallization protocols developed by LUMINEU \cite{Berge:2014,Grigorieva:2017,Armengaud:2017}, 32 LMO cubic crystals (45 mm side) were produced from molybdenum enriched to $\sim$98\%; the same crystal size has been adopted by CUPID. The total crystal mass is 2.4 times larger than that used in CUPID-Mo. 
Despite the promising results achieved with small-scale CROSS prototypes (4 cm$^3$) \cite{Bandac:2020,Bandac:2021}, the detector performance of large-volume (90 cm$^3$) LMO crystals with surface metal coating was found to be degraded \cite{Khalife:2021} and further investigation is needed for possible application of this technology in future cryogenic $0\nu2\beta$ search experiments, such as the CUPID extension CUPID-1T \cite{CUPID1T:2022}. 
Therefore, the CROSS Collaboration decided to abandon the surface coating of LMOs and to pair each crystal with a cryogenic light detector (LD) to detect LMO scintillation, providing particle identification \cite{Poda:2021}. In order to enhance LD performance, a thermal signal amplification based on the Neganov-Trofimov-Luke (NTL) effect \cite{Neganov:1985,Luke:1988} will be exploited, allowing us to demonstrate this technology for CUPID. 
A dedicated low-background cryogenic facility was installed at the LSC in 2019 and has been actively used in the CROSS R\&D to date \cite{Armatol:2021b,CROSS_Magnetic_dampers:2023}. In addition to massive passive shielding of the facility, a muon veto has been installed around the cryostat \cite{CROSS_MuonVeto:2026}. A mechanical detector structure with a very low amount of passive materials, namely copper and polylactide (PLA), has been developed and tested for the CROSS experiment \cite{CROSSdetectorStructure:2024}. 
The CROSS detector array was constructed and installed at the LSC in mid-2025. The commissioning of the CROSS experiment was carried out between November 2025 and March 2026, and regular physics data taking has been ongoing since April 2026.

In the present paper, we describe the construction and underground installation of the CROSS detector (Sect. \ref{sec:detector}), dedicated Monte Carlo simulations of background expected from detector components (Sect. \ref{sec:background}), and an estimate of the experimental sensitivity to the $^{100}$Mo $0\nu2\beta$ decay according to background projections (Sect. \ref{sec:sensitivity}).

%%=======================================================
\section{CROSS detector design, construction and installation}
\label{sec:detector}

The CROSS experiment operates an array of 42 modules of cryogenic calorimeters with dual heat-light readout. This section describes the CROSS detector components such as crystals (Sect. \ref{sec:crystals}), light detectors (Sect. \ref{sec:LDs}), sensors (Sect. \ref{sec:sensors}) and the mechanical structure together with the assembly procedure (Sect. \ref{sec:assembly}).

%---------------------------------------------------------
\subsection{Crystals}
\label{sec:crystals}

\begin{figure}
    \centering
    \includegraphics[width=0.4\linewidth]{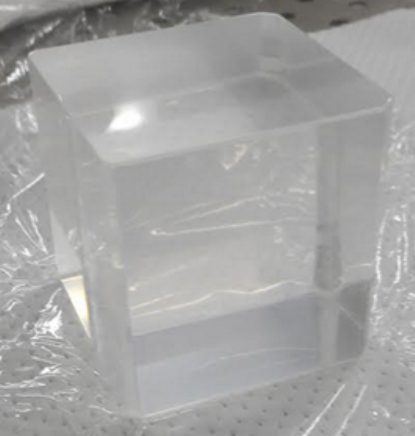}
    \includegraphics[width=0.45\linewidth]{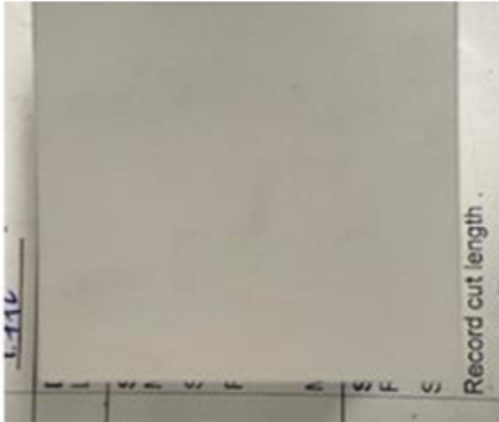}
    \caption{Images of Li$_{2}$$^{100}$MoO$_4$ (left) and $^{130}$TeO$_{2}$ (right) crystals. 
       }
    \label{fig:crystals}
\end{figure}

The CROSS detector array contains 42 cubic crystals (45~mm side) based on $\sim$0.28~kg LMOs and $\sim$0.56~kg tellurium dioxide (TeO$_{2}$, TeO), which are specially developed from radiopure materials. They are grouped as follows:  
\begin{itemize}

    \item 32 Li$_{2}$$^{100}$MoO$_4$ crystals grown by a double crystallization process with the low-thermal-gradient Czochralski (Cz) technique from molybdenum enriched in $^{100}$Mo to 97.7(3)\% \cite{Armatol:2021}. The total mass of Li$_{2}$$^{100}$MoO$_4$ crystals is 8.93~kg, which corresponds to 4.9~kg of $^{100}$Mo. The purification and crystallization technologies developed by LUMINEU \cite{Berge:2014,Grigorieva:2017,Armengaud:2017} and adopted for CUPID-Mo \cite{Armengaud:2019} were exploited, allowing us to obtain high optical quality, transparent scintillating crystals, as illustrated in Fig. \ref{fig:crystals} (left).     
    Approximately half of the crystals have been tested as scintillating thermal detectors that show high performance (energy resolution of $\sim$6 keV FWHM at 2615 keV $\gamma$ peak) and acceptable light output ($\sim$0.3 keV detected scintillation signal per 1 MeV energy released in the crystal as heat) \cite{CrossCupidTower:2023a,CROSSdetectorStructure:2024,CUPID_gdpt:2025}. The ratio between the energy detected in the light detector and that detected in the large crystal is defined LHR (light-to-heat ratio) and is measured in keV/MeV.
    The internal activity of $^{228}$Th and $^{226}$Ra is demonstrated to be below 1 $\mu$Bq/kg \cite{CrossCupidTower:2023a}. 
    
    \item 6 $^{130}$TeO$_{2}$ crystals developed within the CROSS project using highly purified tellurium powder enriched in $^{130}$Te at 91.4\% and the Cz crystal growth method \cite{CROSS_enriched_TeO:2024}. All surfaces of the produced cubic crystals were grounded, as seen in Fig. \ref{fig:crystals} (right). The total mass of the $^{130}$TeO$_{2}$ crystals is 3.333~kg, corresponding to 2.4~kg of $^{130}$Te. Half of the samples have already been operated as thermal detectors in the CROSS setup with promising results, in particular, the internal activity of $^{228}$Th and $^{226}$Ra is found to be less than 10~$\mu$Bq/kg \cite{CROSS_enriched_TeO:2024}.
    
    \item 2 LMO samples produced from molybdenum depleted in $^{100}$Mo, following the same protocol as for the $^{100}$Mo-enriched samples. These crystals, being operated as thermal detectors, demonstrate similar performance, radiopurity, and scintillation properties as natural and $^{100}$Mo-enriched crystals \cite{CROSSdeplLMO:2023}. These two samples will be used for background studies in the CROSS experiment. 
    
    \item 2 LMO crystals produced from molybdenum with natural abundance by the SIMaP laboratory (Saint-Martin de'H\`eres, France) and the SICCAS company (Shangai, China) using the Cz and Bridgman crystal growth methods respectively, as part of R\&D activities for CUPID. 
    
\end{itemize}

%---------------------------------------------------------
\subsection{Light detectors}
\label{sec:LDs}

\begin{figure}
    \centering
    \includegraphics[width=\linewidth]{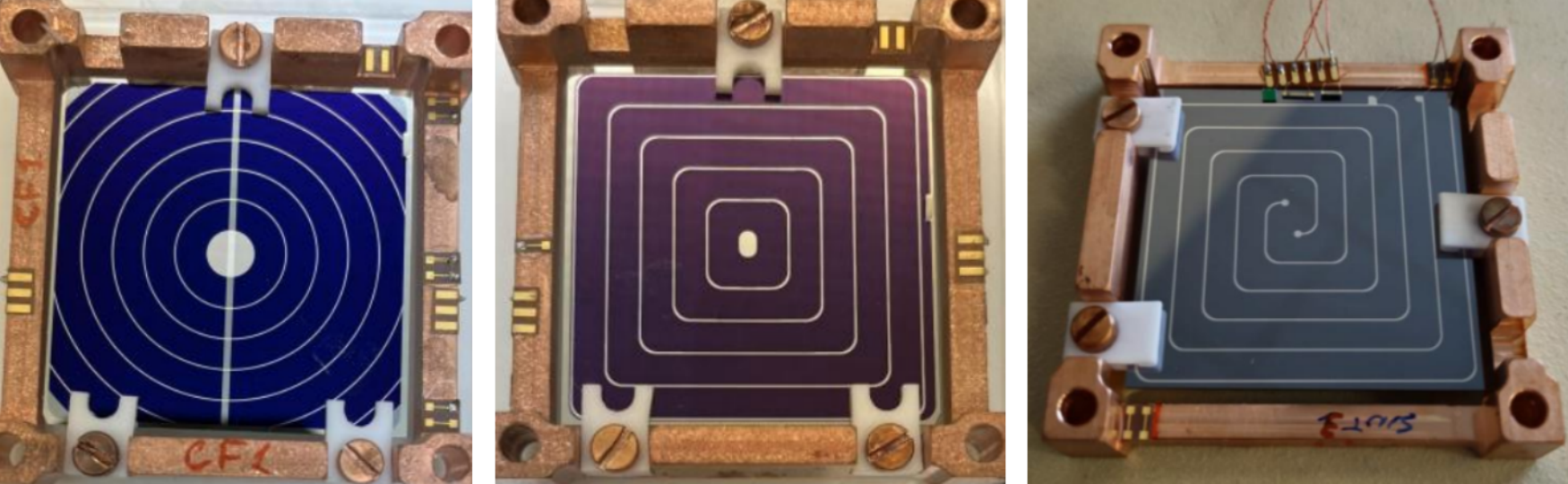}
    \caption{Examples of the LDs produced with each type of mask for the Al electrodes. From left to right: circular, square and spiral electrode shape.}
    \label{fig:LDs}
\end{figure}

Each crystal in the CROSS experiment is coupled to a 45$\times$45~mm$^2$ cryogenic light detector (LD) made of high-purity high-resistivity Ge (by the company Umicore, Belgium) or Si (by the company TOPSIL, Denmark) wafers with thicknesses of 0.30 and 0.65 mm, respectively. In order to enhance the light-assisted particle identification capability and the efficiency of pulse shape discrimination of pile-up events \cite{CROSSpileup:2023}, each LD has Al electrodes evaporated on one surface that allows voltage-driven amplification of the thermal signal, exploiting the Neganov-Trofimov-Luke effect (NTL-LD devices)~\cite{Novati:2019}. Following this technology, 42 square devices were produced, with three Al electrode designs (shown in Fig.~\ref{fig:LDs}), which are grouped as follows:
\begin{itemize}

    \item a) 12 Ge NTL-LDs with circular concentric electrodes; 
    
    \item b) 1 Ge NTL-LD with semi-square concentric electrodes;
    
    \item c) 13 Ge NTL-LDs with double-spiral electrodes; 
    
    \item d) 16 Si NTL-LDs with double-spiral electrodes. 
    
\end{itemize}
The circular electrode geometry of the square-shaped LD provides an electric-field area coverage of $\sim$56\%, while the semi-square and double-spiral designs allow us to increase the coverage to $\sim$100\%. In previous tests of these devices, we observed that the signal-to-noise ratio in light signals scales linearly with the covered area with good approximation.

To improve light collection efficiency, a 70-nm SiO layer, acting as an anti-reflective coating, is deposited on both sides of the Ge wafers; 
the exception is a device with semi-square concentric electrodes having the anti-reflective layer only on the electrode's side. 
In contrast to Ge wafers, Si samples are not coated (an R\&D on anti-reflective coating of Si wafers is ongoing in view of CUPID).

Square-shaped Ge LDs with circular Al electrodes have recently been validated for CROSS with a 10-crystal prototype tower operated at LSC \cite{CROSS_Run9:2025,CROSS_enriched_TeO:2024}. In particular, applying 80 V bias to Al electrodes, the average gain of the LD signal-to-noise ratio is around 9, achieving an RMS baseline noise of $\sim$ 10 eV. Thanks to the operation of LDs at comparatively ``warm'' working points (with resistance of NTDs of $\sim$ 0.9 M$\Omega$), a comparatively fast response has been achieved (0.54 ms on average) .

%---------------------------------------------------------
\subsection{Sensors}
\label{sec:sensors}

Both the cubic crystals and the wafers of the CROSS detectors are instrumented with neutron transmutation-doped (NTD) Ge thermistors as thermal sensors~\cite{Haller:1994}. These thermistors were produced from Ge wafers with similar neutron irradiation parameters. The NTDs used for the crystals are 3~$\times$~3~$\times$~1~mm$^3$, while those glued to the wafers have different geometries. Most of them have a shape of 3~$\times$~0.7~$\times$~1~mm$^3$ or 4~$\times$~0.5~$\times$~0.5~mm$^3$. Several batches of NTDs were glued to the wafers: 21~NTDs were previously used in the LUMINEU project \cite{Armengaud:2017,Poda:2017a}, 9~NTDs were used in the CUPID-Mo experiment \cite{Armengaud:2020a} and the others (those with 4 mm as a longer dimension) were part of a new thermistor production, based on the Ge wafer irradiation campaign performed within the framework of LUMINEU project \cite{LUMINEU_NTD:2016}. In every configuration, the bias current flows along the thermistor's long dimension during operation.

\begin{figure}
    \centering
    \includegraphics[width=0.9\linewidth]{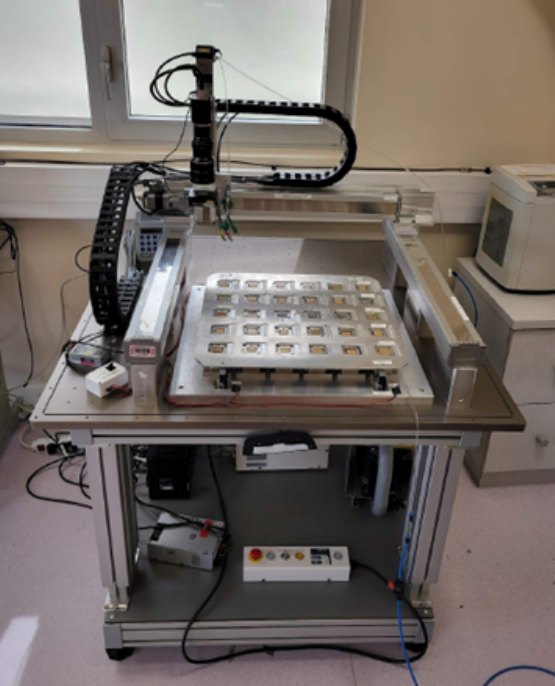}
    \caption{Photograph of the robotic gluing station, developed at CEA within the CUPID project, for a simultaneous NTD sensor gluing on 30 crystals.}
    \label{fig:Robot_gluing}
\end{figure}

\begin{figure}
    \centering
    \includegraphics[width=1.0\linewidth]{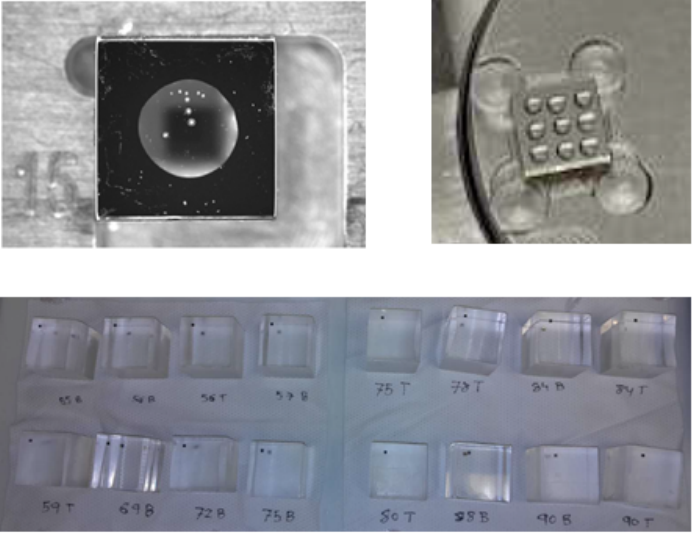}
    \caption{Photograph of a typical spot of the UV-cure glue (top left), 9 spots of bi-component epoxy (top right), and a batch of LMO crystals with glued NTD and heater (bottom).}
    \label{fig:Gluing_examples}
\end{figure}

The NTDs were glued to the crystals using two different coupling materials and with two different procedures. This allows us to compare the detector responses for each configuration. The coupling materials used were UV-curable glue (PERMABOND\textregistered~UV620) and bi-component epoxy (Araldite\textregistered ~Rapid), both of which are well-tested for cryogenic applications and have been demonstrated to have acceptable radiopurity. 

The performance of thermal detector strongly depends on the quality of the gluing. 
In the case of the cubic crystals, previous tests with these bolometers have shown that optimal results are obtained when the NTD is connected to the absorber through separated glue spots when Araldite\textregistered ~epoxy is used. This reduces the thermal stress between the involved materials as a result of their different thermal contractions. Thus, a gap of $\approx$50~$\mu$m is kept between the NTDs and the crystals. The UV-curable glue, being much less viscous than Araldite\textregistered ~epoxy, is applied as a thin layer covering the entire NTD surface. The two gluing procedures follow a similar technique; the difference is in the holding system, which can be done by hand using a dedicated tool (similar to the one used in CUPID-Mo \cite{Armengaud:2020a}) or automatically using a special robotic system developed for CUPID, shown in Fig. \ref{fig:Robot_gluing}. In both cases, the NTDs are held and can be moved along the vertical axis and fixed at any level. At the same time, the robotic system allows us to perform the gluing much faster and with a more precisely controlled amount of glue. 
The robot was used for 26 LMOs, while thermistors of the other 10 samples of this compound were glued manually; the UV-curable glue was used in both approaches. The estimated mass of glue for robotic and manual methods is around 130 $\mu$g and 230 $\mu$g, respectively; a similar quantity of glue was used in manual gluing for CROSS \cite{CROSSdetectorStructure:2024,CROSS_Run9:2025} and CUPID \cite{CUPID_gdpt:2025} prototypes. An image of a UV-curable glue droplet before compression into a thin layer is shown in Fig. \ref{fig:Gluing_examples} (top left). 
All NTD thermistors of TeO crystals were also manually glued using a drop of the UV-curable glue for the half of samples and 9 spots of Araldite\textregistered~Rapid bi-component epoxy for the another half. An image of a 9-spot epoxy glue pattern is shown in Fig. \ref{fig:Gluing_examples} (top right).

In addition to the thermistor, we also coupled a silicon-based resistive chip to each crystal~\cite{Andreotti:2012} using a single spot of Araldite\textregistered~Rapid glue, as illustrated in Fig. \ref{fig:Gluing_examples} (bottom). This chip acts as a heater which allows us to inject fixed amounts of heat and generate thermal pulses. Since they have constant energy, the resulting pulses can be used as a reference in the offline analysis, allowing correction for signal gain changes due to temperature drifts of the bolometers \cite{Alessandrello:1998}. 

For the gluing of NTDs to the Ge and Si wafers, we used 3~spots of Araldite\textregistered~Rapid epoxy, compressed into a layer. In order to prevent an electrical contact between the wafer and the NTD (that we suspect may generate leakage currents), a protection made of hardened UV-curable glue (PERMABOND\textregistered~UV645) was deposited between the wafer and the sensor before the epoxy gluing. In the case of the LUMINEU NTDs, a veil of UV-curable glue was used, while for the other batches of thermistors a set of 6 spots provided the necessary separation. Similarly, suitable transition pieces were attached using a small amount of Araldite\textregistered~Rapid epoxy, with a protective layer of UV-curable glue (PERMABOND\textregistered~UV645) preventing direct contact. These elements consist of small Si parallelepipeds coated with a continuous Al film extending over the top surface and one lateral face. This arrangement enables wire bonding between the electrodes and the top surface, as well as between the lateral face and the pads connected to the cryostat wiring.

%---------------------------------------------------------
\subsection{Detector structure and assembly}
\label{sec:assembly}

The CROSS detector represents a modular structure~\cite{CROSSdetectorStructure:2024} developed according to the strict requirements of a low-background experiment, particularly using a small amount of radiopure passive elements near the detectors. Therefore, the Cu-to-LMO mass ratio is kept at a very low level (6\%) to minimize the radioactivity coming from the bulk and surface of close elements. Although the light emitted by the crystals is low, no reflective foil is used around the crystals. This avoids the presence of surface radioactivity and also improves the coincidences between detectors.

\begin{figure}
    \centering
    \includegraphics[width=1.0\linewidth]{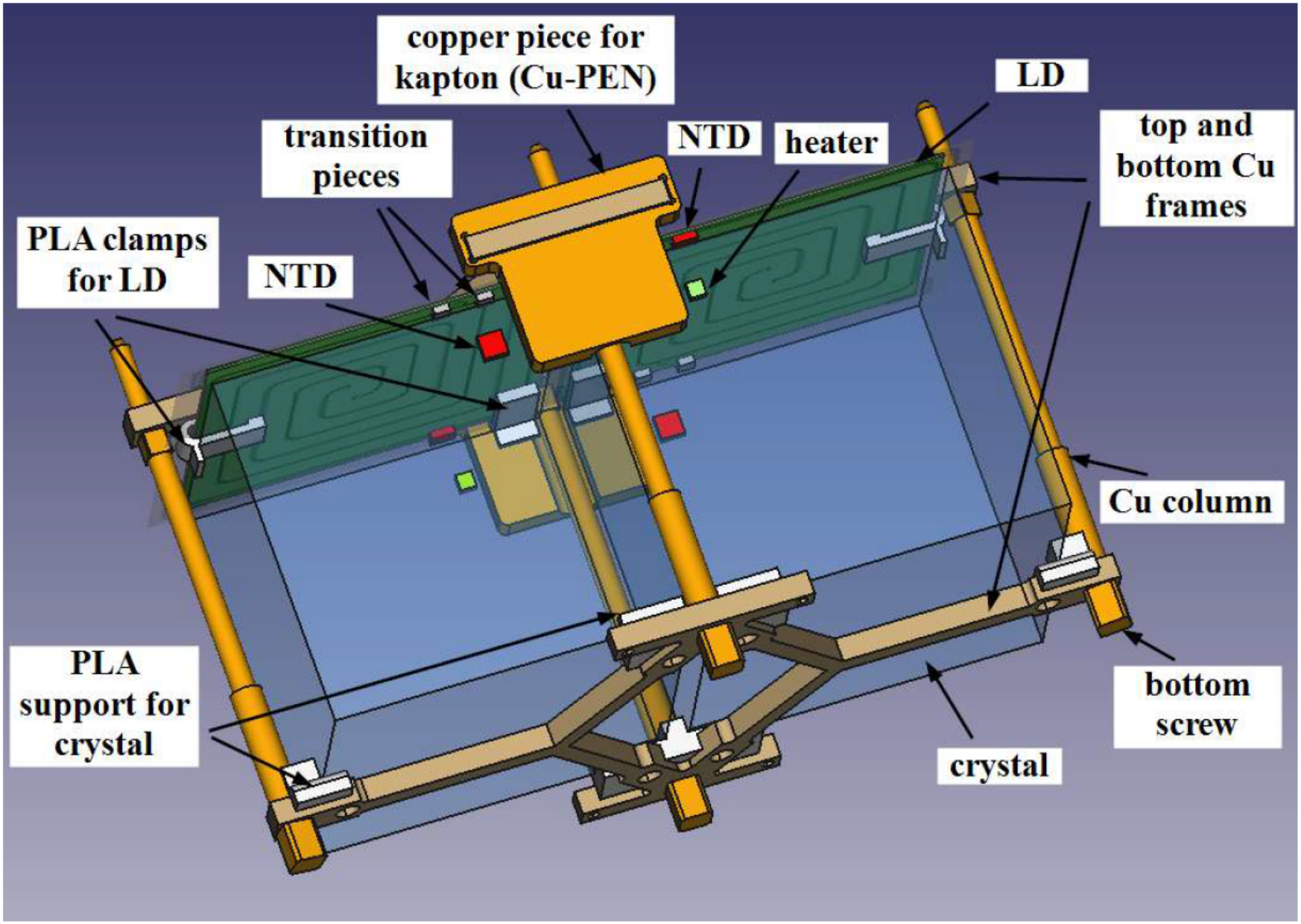}
    \caption{Rendering of the mechanical structure of a two-crystal module developed for the CROSS experiment (see details in \cite{CROSSdetectorStructure:2024}).}
    \label{fig:Detector_Structure}
\end{figure}

\begin{figure*}[!ht]
    \centering
    \includegraphics[width=0.80\linewidth]{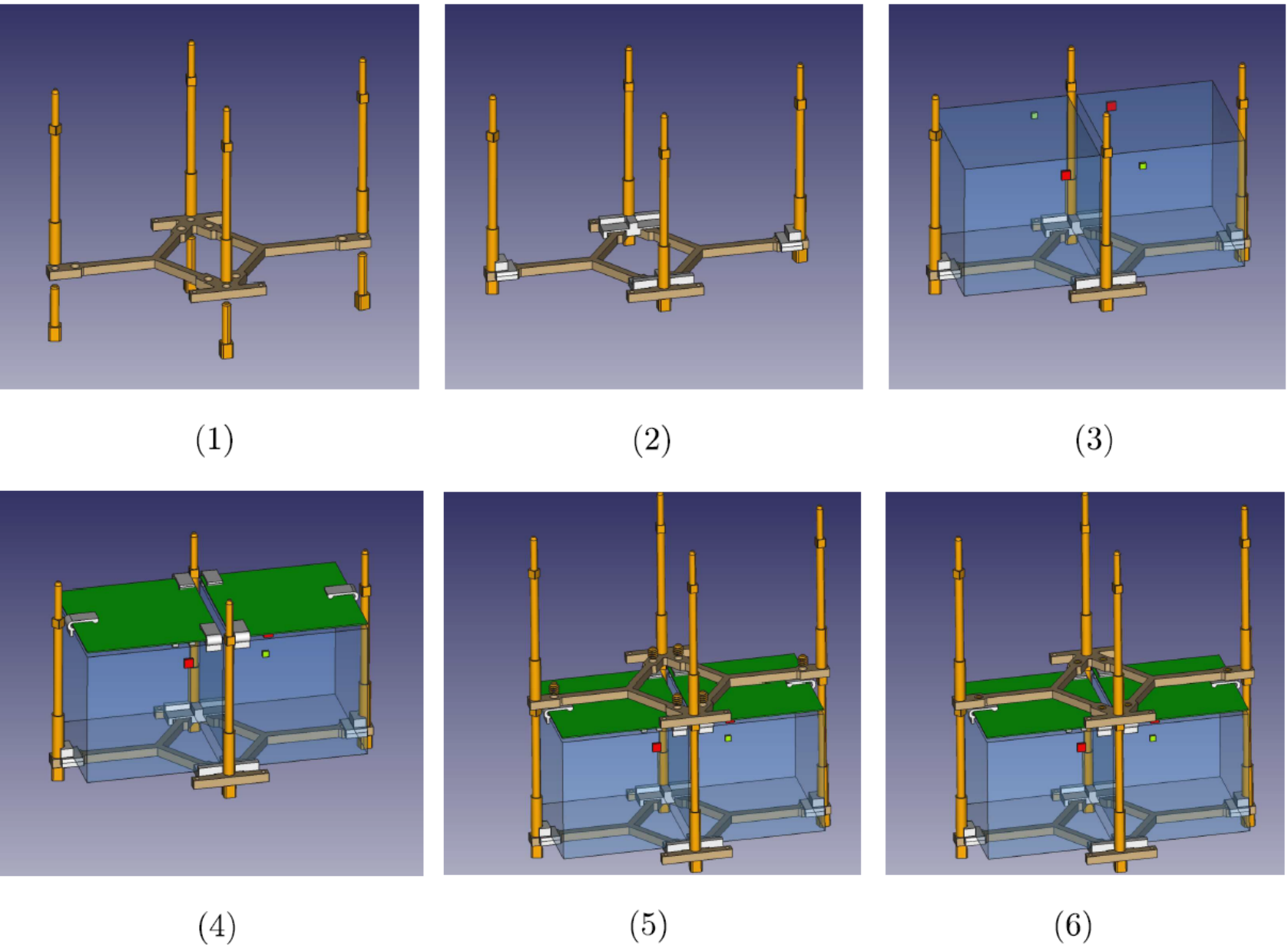}
    \caption{Main steps of the CROSS detector array construction: 
    (1) Assembly of the Cu structure; (2) Placement of the PLA supporting pieces; (3) Installation of two cubic crystals; (4) Placement of two LDs on top of the crystals; (5) Placement of the next floor Cu structure; (6) Screwing nuts to apply a force on LDs via the PLA clamps.}
    \label{fig:Detector_Assembly}
\end{figure*}

Figure~\ref{fig:Detector_Structure} shows the design of a single 2-crystal module of the CROSS array, while Fig.~\ref{fig:Detector_Assembly} illustrates the main steps of the detector assembly. 
Two cubic crystals are mounted on the Cu~frame and thermally decoupled by 3D-printed PLA pieces, which also confine the position of the crystals and the wafers laterally. The wafers are placed on top of the crystals and held using three 3D-printed PLA clamps, which apply a force towards the crystal with a small Cu~screw. An identical Cu~frame is placed above, and the construction is fixed using four Cu~columns and nuts. 
These structures holding two crystals and two wafers were recursively mounted on top of each other (see Fig.~\ref{fig:Detector_Assembly}) to build a tower.

\begin{figure}
    \centering
    \includegraphics[width=1.0\linewidth]{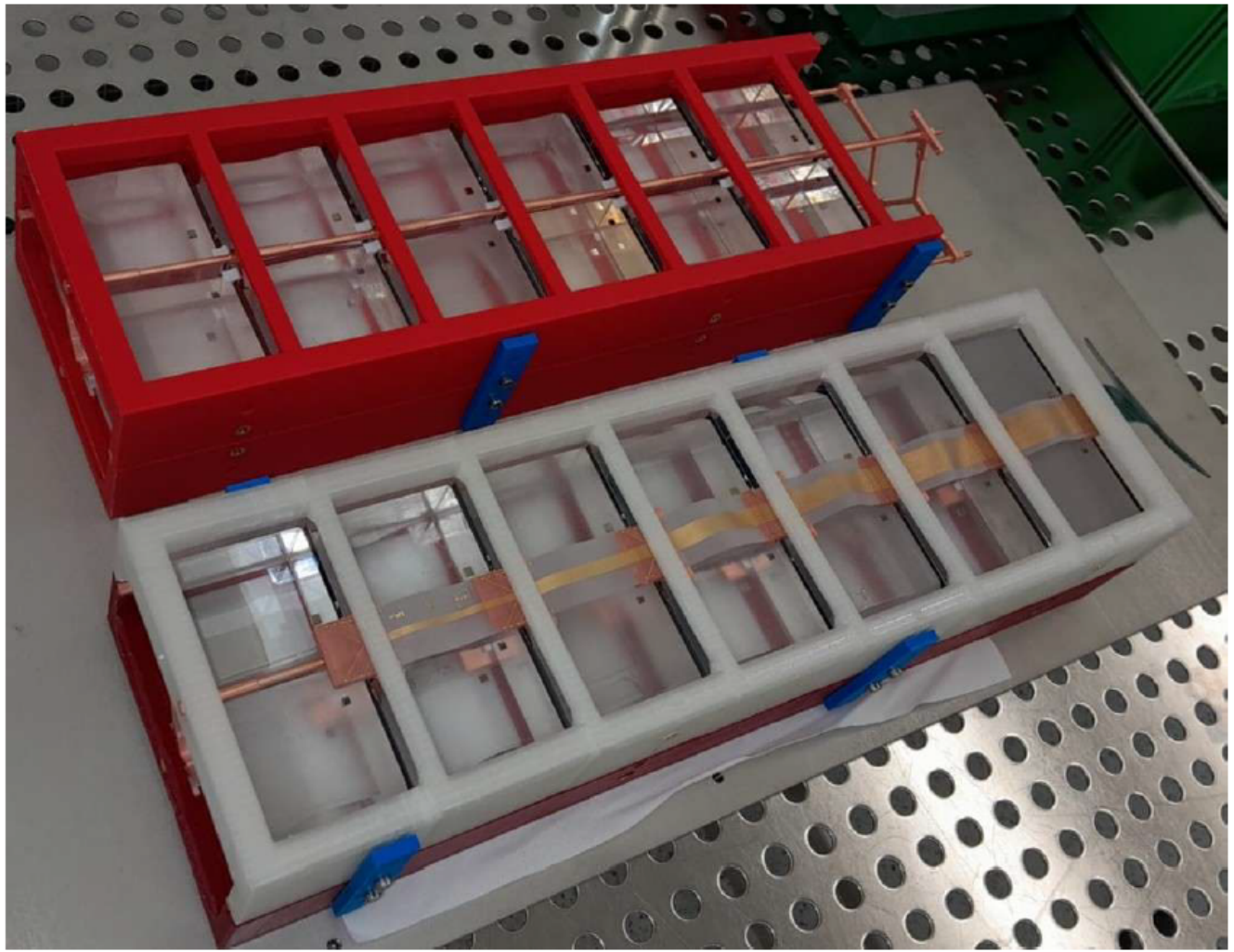}
    \caption{Two CROSS towers inside an individual 3D-printed skeleton made of PLA.}
    \label{fig:CROSS_exoskeleton}
\end{figure}

The CROSS detector array assembly, consisting of three 7-floor towers, was performed in a cleanroom (class ISO5 at worst) at IJCLab. 
All detector components were cleaned before mounting to minimize the possibility of surface contamination. The copper elements were cleaned with radiopure soap (Micro-90\textregistered) and etched with citric acid in an ultrasonic bath, and the PLA pieces were cleaned with ethanol. 
In order to facilitate the handling of each tower during the construction and further manipulations (like wiring and bonding), we used 3D-printed PLA exo-skeletons shown in Fig.~\ref{fig:CROSS_exoskeleton}. 

The wiring scheme designed for the demonstrator accommodates the 84~readout channels required for the experiment, as well as the connections for the heaters and the electrodes of the wafers. 
Initially, the wiring of each CROSS towers was implemented using two flexible Cu-PEN (copper on a polyethylene naphthalate substrate) strips with 64 contacts each glued with Araldite\textregistered~Rapid epoxy (see on one tower of Fig. ~\ref{fig:CROSS_exoskeleton}). Unfortunately, after the completion of the full array wiring, we found that around 25\% of contacts were lost and it forced us to change the wiring design by adopting the CUPID-Mo approach \cite{Armengaud:2020a}. 
Thus, the sensors were electrically connected to Kapton pads glued onto the detector holder, which also provided a weak thermal link to the heat bath. Gold wires were used to bond the NTDs and heaters, while aluminum wires were employed for the transition pieces. Insulated twisted copper wires were soldered to the holder Kapton pads and then connected, via additional Kapton pads on top of the towers. The heaters are parallelized across the semi-tower.

\begin{figure*}[!ht]
    \centering
    \includegraphics[width=0.8\linewidth]{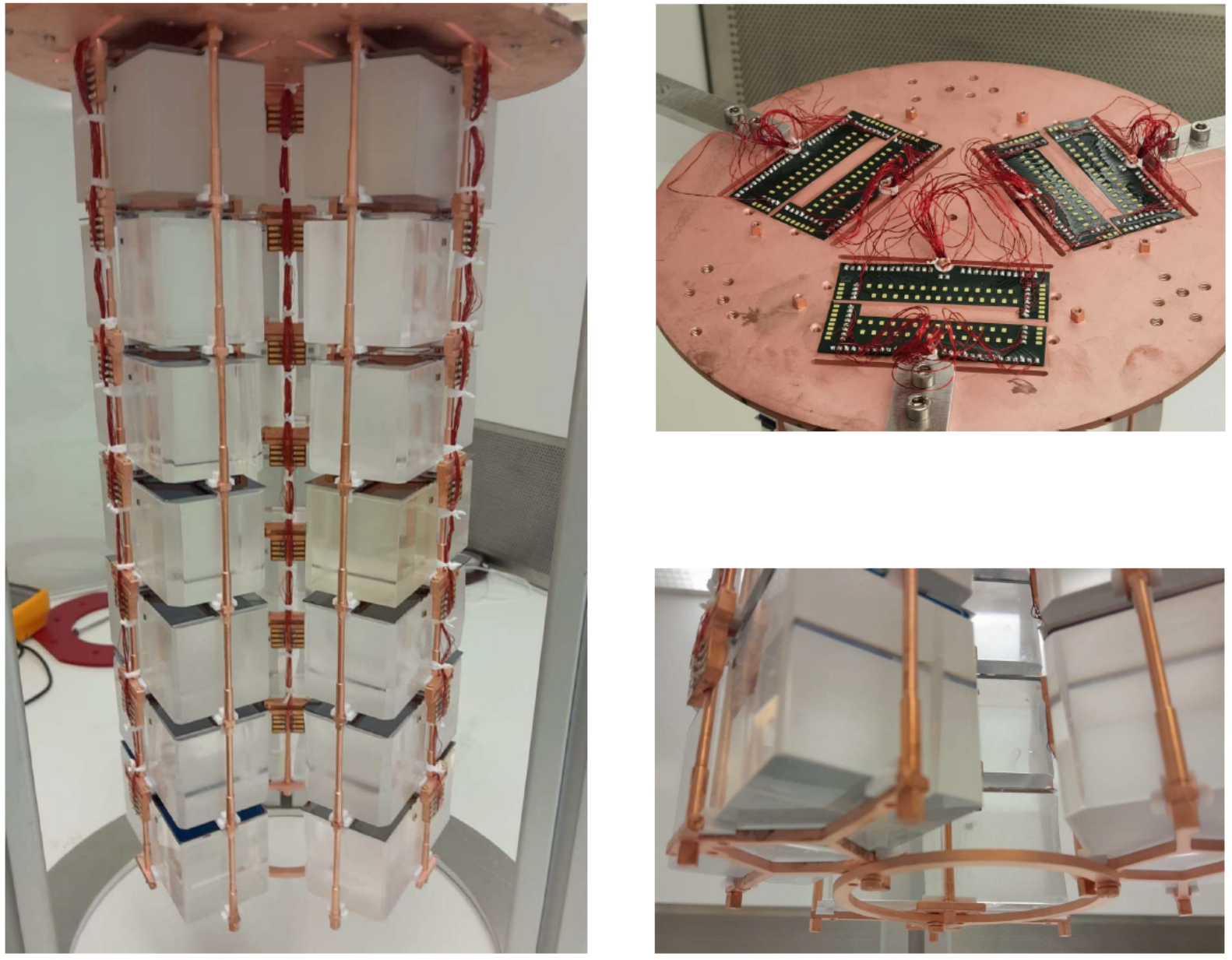}
    \caption{Photograph of the CROSS detector array (left) together with top and bottom view showing respectively the upper part of wiring (right top) and a Cu ring for confinement of the towers (right bottom).}
    \label{fig:CROSS_array}
\end{figure*}

\begin{figure}
    \centering
    \includegraphics[width=0.8\linewidth]{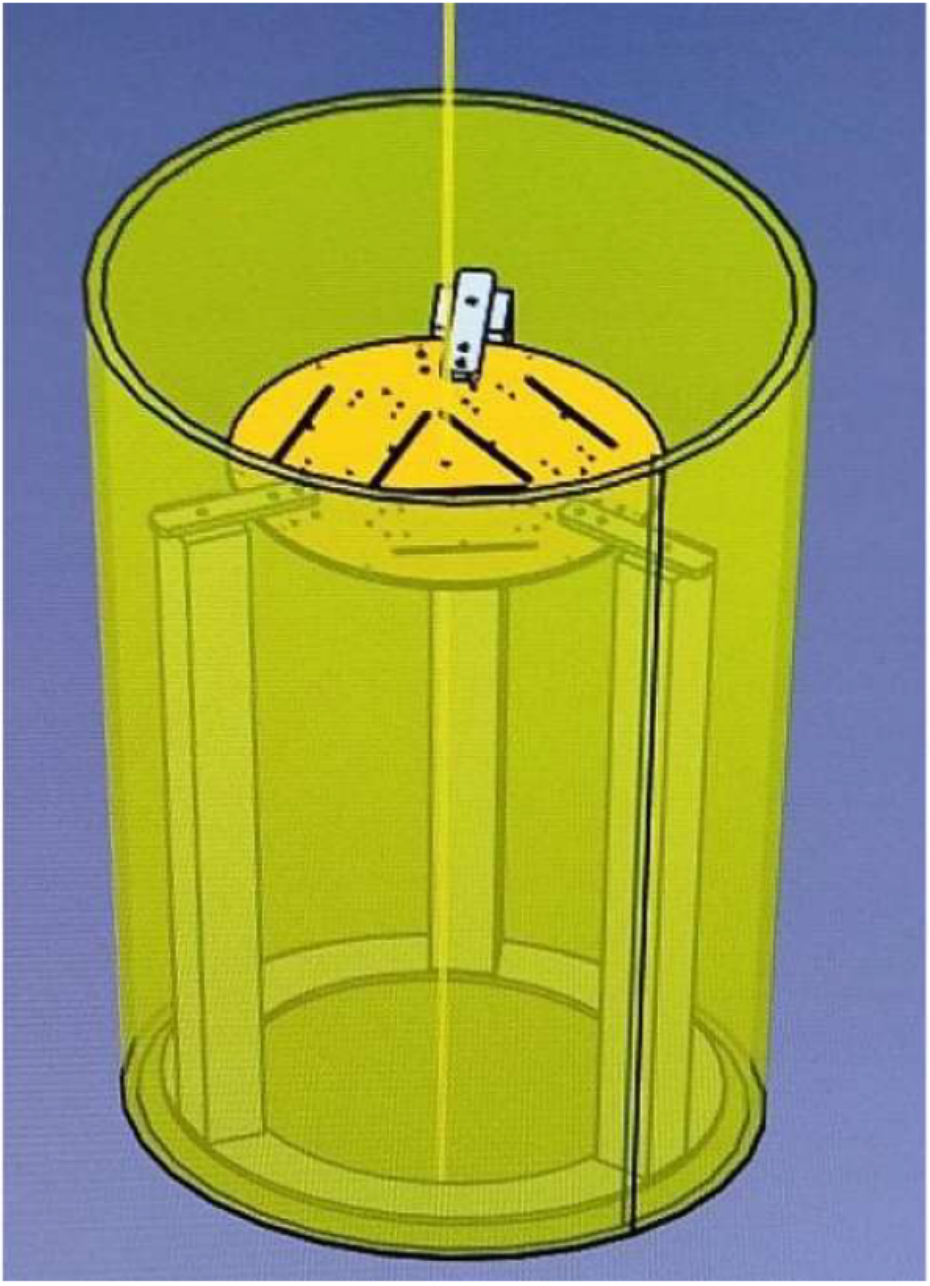}
    \caption{Drawing of a vacuum-tight container with a supporting structure to suspend the CROSS array.}
    \label{fig:CROSS_container_transportation}
\end{figure}

The three wired towers were then attached to a copper plate (Fig. \ref{fig:CROSS_array}) and suspended using a 3-column structure with dampers (Fig. \ref{fig:CROSS_container_transportation}) to mitigate vibrations during the array transportation. The structure with the CROSS array was placed inside a vacuum-tight container (Fig. \ref{fig:CROSS_container_transportation}) flushed with nitrogen gas. As an additional precaution, we put silica gel crystals in the container. Such measures are required due to hygroscopic properties of LMO crystals, which should be kept under humidity below 50\%.

%---------------------------------------------------------
\subsection{CROSS array installation in the Canfranc underground laboratory}
\label{sec:installation}

The CROSS detector array was transported from Orsay (France) to Canfranc underground laboratory (Spain) and installed inside a dedicated low-background cryogenic setup (C2U)~\cite{Olivieri:2020} in the end of July 2025. The underground location of the laboratory provides the suppression of cosmic muon flux corresponding to 2450~m of water equivalent (m~w.e.). 
The C2U facility is equipped with a dilution refrigerator by CryoConcept (France) and uses a pulse tube (Cryomech\textregistered PT415) to cool down to 4~K. The cryostat is also assisted by Ultra-Quiet Technology\texttrademark (UQT)~\cite{UQT} to mechanically decouple the pulse tube from the dilution unit, thus reducing vibrations that can affect detector performance~\cite{Olivieri:2017}. To further mitigate residual vibrations, a detector suspension system is implemented ~\cite{CROSS_Magnetic_dampers:2023}. It consists of three springs installed at the 1~K stage provided with magnetic dampers. The natural frequency of this suspension system is 3.5~Hz.

The passive shielding of the set-up against environmental radiation consists of lead (25~cm thickness minimum) which protects the experimental volume laterally and from the bottom. Furthermore, a 13-cm-thick Pb and Cu internal shielding and Pb layers between thermal screens are installed above the experimental volume to shield it from the top part of the cryostat. Finally, an anti-radon box is installed around the lead shield and is continuously flushed with deradonized air ($\sim$~1~mBq/m$^3$ of Rn).

A muon veto system surrounds the shielding. It is constructed from 174~individual plastic scintillator modules, divided in 4 lateral sectors (112 modules), 4 bottom sectors (60 modules) and a single top sector (2 modules). Given the relatively high muon flux in Hall B of LSC ($\sim$20~$\mu$/m$^2$/h)~\cite{Trzaska:2019}, it becomes essential to apply a strong muon rejection: any event in coincidence between an LD and any of the veto modules within a time window of 2~ms is excluded from the analysis. This allows us to reject $\sim$80\% of muon-related events in the region of interest (ROI) with a $\sim$18\% dead time \cite{CROSS_MuonVeto:2026}.

The data acquisition system of the CROSS facility is based on room-temperature low-noise voltage-sensitive DC electronics implemented with 24-bit~ADC on 12-channel cards~\cite{Carniti:2020,Carniti:2023}. The output voltage of the thermistors is acquired as a continuous data stream with a defined sampling frequency and stored on a hard disk. Data processing is done offline using Gatti-Manfredi optimum filtering \cite{Gatti:1986}. 
A removable $^{232}$Th $\gamma$ source, which can be inserted inside the Pb shield, is used to calibrate thermal detectors in the C2U setup. 

Due to an issue with a water cooling system (chiller) supplying the C2U facility, the cooling down of the CROSS detector array was realized in October 2025. 
The commissioning of the CROSS detectors was carried out during the period from November 2025 to March 2026, and regular physics data taking has been ongoing since April 2026. 
The results of the CROSS commissioning will be reported in a dedicated article.

%%=======================================================
\section{Background model of the CROSS experiment}
\label{sec:background}

To explore the background expected in the CROSS experiment, Monte Carlo (MC) simulations were performed using the Geant4~11.1 toolkit~\cite{Agostinelli:2003}. In this section, we present the geometry built (Sect.~\ref{sec:Geometry}), the detector response model implemented in such simulations (Sect.~\ref{sec:Response}), the considered background sources (Sect.~\ref{sec:Sources}), the event selection (Sect.~\ref{sec:EventSelection}), and the projected background index (BI) in the region of interest for $^{100}$Mo $0\nu2\beta$ decay (Sect.~\ref{sec:BI}).

%---------------------------------------------------------
\subsection{Geometry of Monte Carlo simulations}
\label{sec:Geometry}

\begin{figure*}[!ht]
  \centering
  \includegraphics[height=0.39\textheight]{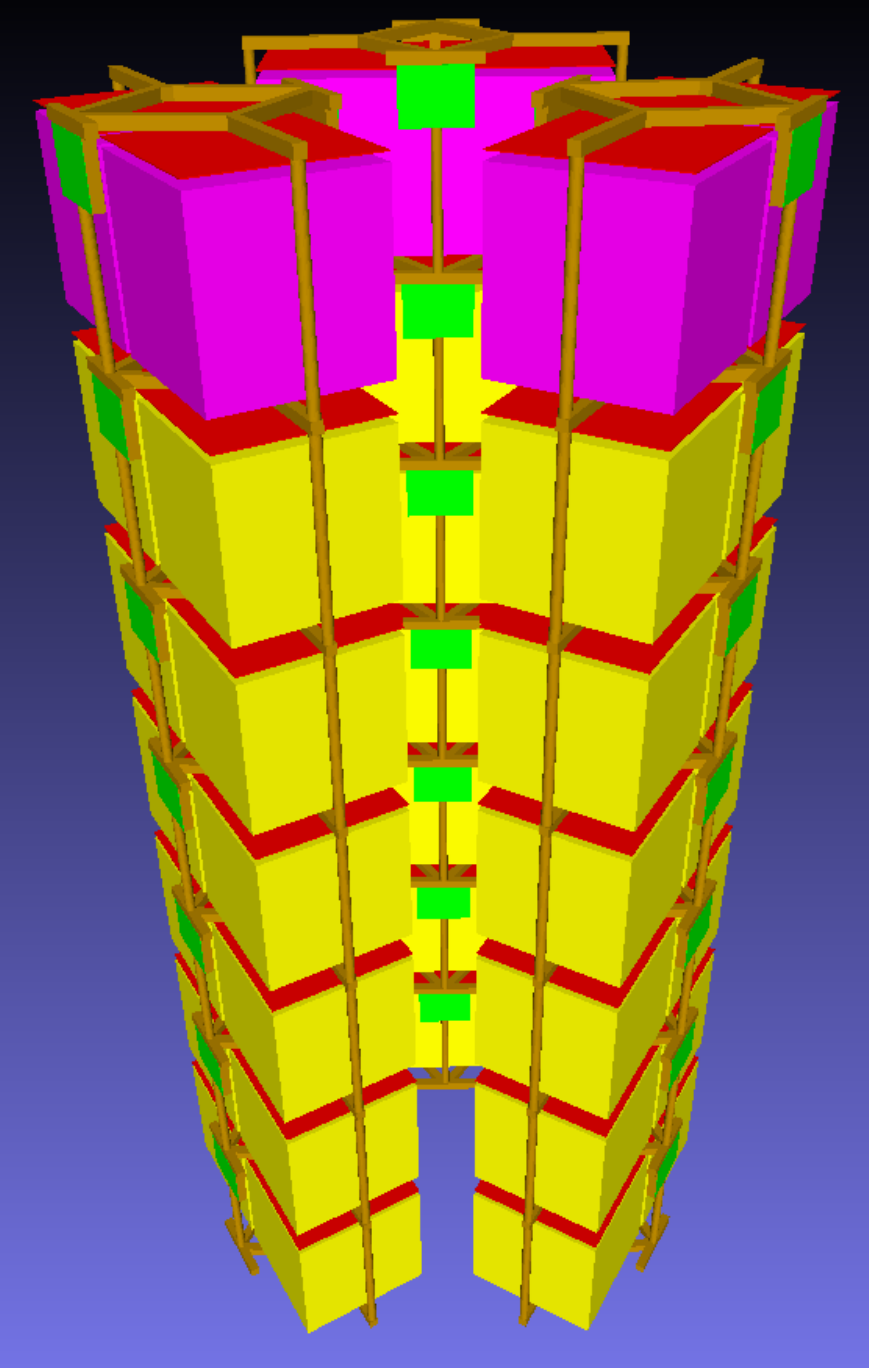}
  \includegraphics[height=0.39\textheight]{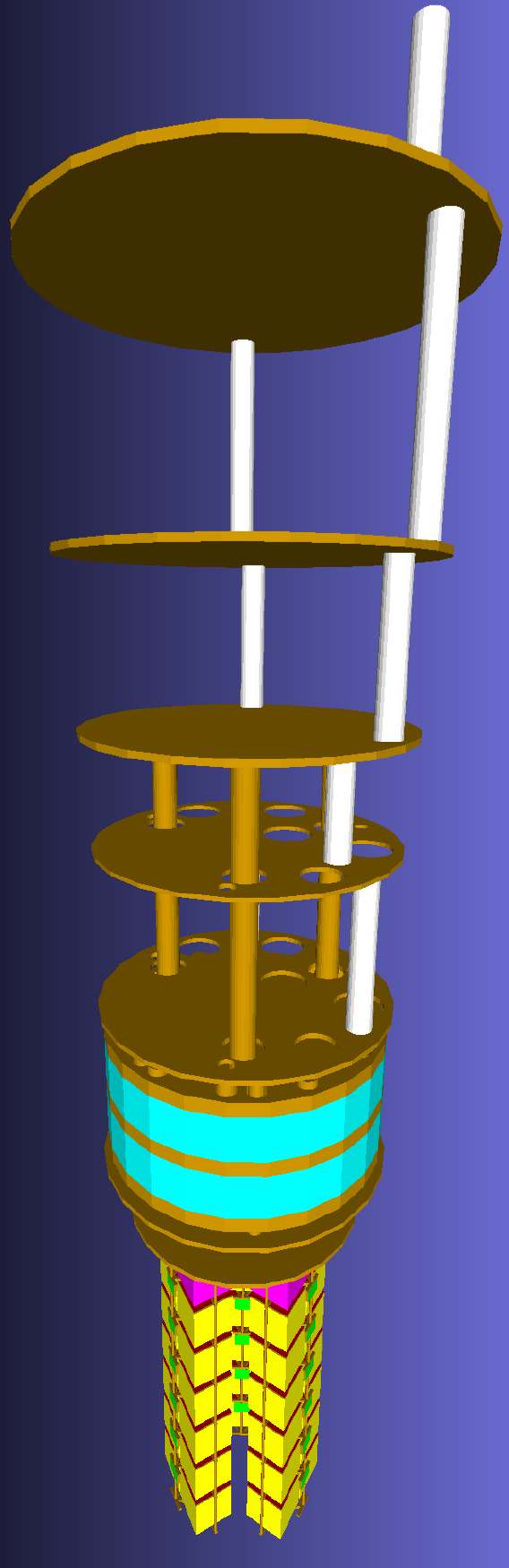}
  \includegraphics[height=0.39\textheight]{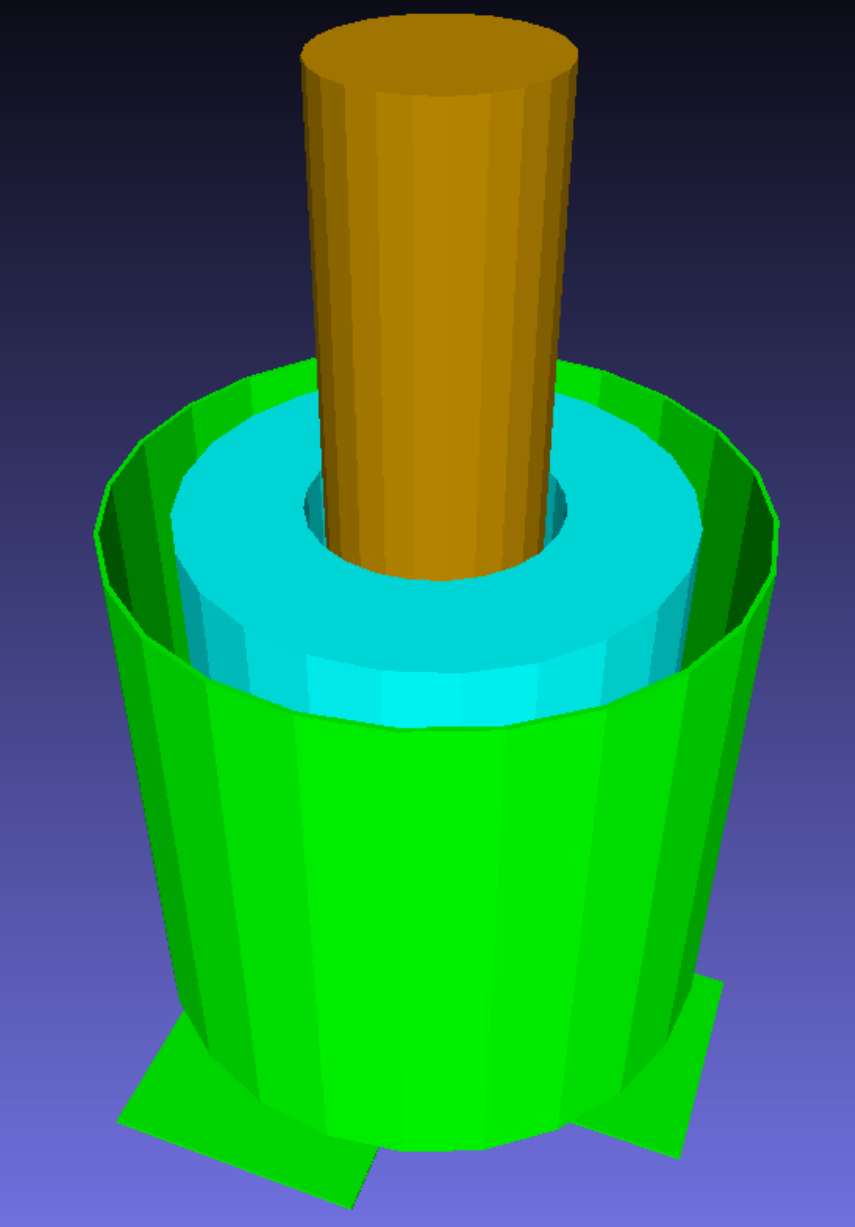}
  \caption{Rendering of the simulation geometry of the CROSS experiment. 
  (Left) The geometry of the detector array showing LMOs (represented as yellow cubes), TeOs (purples), LDs (red), the copper structure (orange) and the Kapton pads (green). 
  (Middle) Rendering of the internal structure of the cryostat: copper is drawn in orange, lead is in cyan and the fiberglass bars is in white. 
  (Right) The geometry from the external parts of the CROSS facility such as: the outer thermal screen of the cryostat (drawn in orange), the lead shield (cyan) and the muon veto (green), but the top sector placed $\sim$3~m above the lateral veto; see details in text. }
  \label{fig:Geometry}
\end{figure*}

A detailed geometry of the CROSS experiment was implemented in the MC simulations. 
The volumes simulated are: the crystals (LMOs and TeOs), the LDs, the copper structure, and the Kapton pads for the signal readout. Some internal components of the cryostat were also simulated: the 6 copper screens and their respective copper plates, the fiberglass bars that connect adjacent copper plates, the internal lead shielding, the copper bars that hold the lead shielding to the 1~K plate, and the copper bars that hold the wires when they pass through the lead shielding. Finally, the volumes external to the cryostat that have been simulated are the external lead shielding and the muon veto. 
Figure~\ref{fig:Geometry} shows the Geant4 rendering of the simulation geometry of the detector array, the cryostat, and the shielding together with the lateral and bottom panels of the muon veto (the full geometry of the CROSS muon veto, including the top sector placed several meters atop, can be found in \cite{CROSS_MuonVeto:2026}).

%\onecolumn
%\nopagebreak

The active volumes of this simulation (those where the energy deposit is stored) are the LMO and TeO crystals, the LDs and the muon veto system. The muon veto is divided into seven sectors (4 lateral, 4 bottom and 1 top), as used in the 
experiment. However, a simplified model of each sector was used, without considering the modular granularity.

%---------------------------------------------------------
\subsection{Detector response model}
\label{sec:Response}

The response of each detector included in the simulation has been modeled, since it is essential to account for the finite energy resolution and response of the detectors. All detectors of the same type (crystals and LD) were assumed to have exactly the same performance as reported in~\cite{CROSS_Run9:2025}, considering the typical performance of the detectors in the optimal working point. It is worth noticing that because of this decision, the detector response model simulates the best working conditions of all detectors of the experiment.

The first implementation in the simulation of the detector response is the time integration window. Once there is a first hit of a particle in a crystal or LD, all energy deposits in each detector are integrated in a time window of 1~ms. This time window is considered because the rise time of the pulses in LDs is on the order of a few hundreds of~ms, so the scintillation or Cherenkov light emitted in the crystals after a particle interaction is triggered by an LD in a time shorter than 1~ms. Therefore, in the simulation, it is considered that any delayed coincidence with a $\Delta t <$~1~ms cannot be identified. In a real scenario, this minimum $\Delta t$ is lower than 1~ms, but this value has been chosen to simulate the worst situation.

All hits that occur with $\Delta t >$~1~ms are considered to be a different event from the previous. In such a case, the first hit that meets this condition is taken as the first energy deposit of a new event, and the $\Delta t$ of the next hits are calculated with respect to that one. This process is repeated until the end of the simulation of the primary event.

For each integrated event, the value of $\Delta t$ of the first hit with respect to the first hit of the previous event is stored. 
This allows us to reject pile-up events when $\Delta t \in [1,x]$~ms. In the experimental data, the range where a pile-up can be identified and therefore rejected depends on the time window, $t_W$, used in the analysis of the data stream. Since in the analysis the maximum of the pulse is set in the middle of the time window, all the events with a $\Delta t < t_W/2$ will fit inside the window and will be identified as pile-ups. Given the decay times of the detectors of a few hundreds of~ms, a typical value for $t_W$ is 1~s. Therefore, all delayed coincidences between 1~ms and 500~ms will be rejected from the analysis.

Once all the energy deposits are integrated into each crystal for a given event, a smearing is applied to account for the detector energy resolution. To do that, the new value of the energy is obtained by taking a random value from a probability density function (PDF). This PDF is defined as a Gaussian with a mean that is the original energy deposited in this detector, $E^0_i$, and with a $FWHM$ given by the energy resolution function 
\begin{equation}
\label{eq:FWHM}
FWHM = a+b\sqrt{E^0_i}.    
\end{equation}

The values of $a$ and $b$ included in the simulation are obtained from the fit of the data from the CROSS detector module described in \cite{CROSS_Run9:2025}; the resulting parameters are 
0.94(18)~keV and 0.100(7)~$\sqrt{\textrm{keV}}$, respectively. Therefore, the energy resolution of the 2615~keV $\gamma$ quanta emitted in $^{208}$Tl decays ($^{232}$Th family) --- the reference calibration point --- is $\sim$6~keV FWHM, while the resolution at the $^{100}$Mo $0\nu2\beta$ ROI (3034~keV) is expected to be around 6.4~keV FWHM.

The scintillation light of the LMOs is also summed to the energy deposited in the LDs considering typical LHR values measured in the CROSS detector structure (i.e., 0.3 keV/MeV for a top LD with respect to the coupled LMO crystal and 2/3 of this value for the bottom one \cite{CROSSdetectorStructure:2024,CROSS_Run9:2025}). Therefore, the total energy deposited in an LD $i$, which is above the LMO $i$ and below the LMO $i+1$ is given by 
\begin{flalign}
\label{eq:ly}
  \nonumber
  &  E_{LD}(i) = E^0_{LD}(i)+10^{-5}\cdot[30\cdot E^{\beta/\gamma}_{LMO}(i) \\  
  &  +20\cdot E^{\beta/\gamma}_{LMO}(i+1) + 5\cdot E^\alpha_{LMO}(i)+3\cdot E^\alpha_{LMO}(i+1)],
\end{flalign}
where $E^0_{LD}$ is the energy deposited in that LD by the simulated particles, $E^{\beta/\gamma}_{LMO}$ is the energy deposited by $\gamma$($\beta$) interactions, $E^{\alpha}_{LMO}$ is the energy deposited by $\alpha$ particles. This scintillation light is summed only in the LDs that are facing LMOs and not in those facing TeOs.

Once the total energy deposited in each LD is calculated, the smearing is applied by the energy resolution in the same way as for the LMOs and TeOs, but using the energy resolution extracted from the approximation of the CROSS prototype data of Ref. \cite{CROSS_Run9:2025} with Eq.~\ref{eq:FWHM}, resulting in $a$ = 0.042(19)~keV and $b$ = 0.31(4)~$\sqrt{\textrm{keV}}$. 
We emphasize that no NTL amplification of the LD thermal signals is considered in simulations; it corresponds to the worst case scenario where no device can be operated in this mode. Taking into account the results of the test of some devices \cite{CROSS_Run9:2025}, we believe that the performance (i.e., energy resolution) of LDs in CROSS will be further enhanced due to the NTL effect.

\begin{figure*}[!ht]
  \centering
  \includegraphics[width=0.49\linewidth]{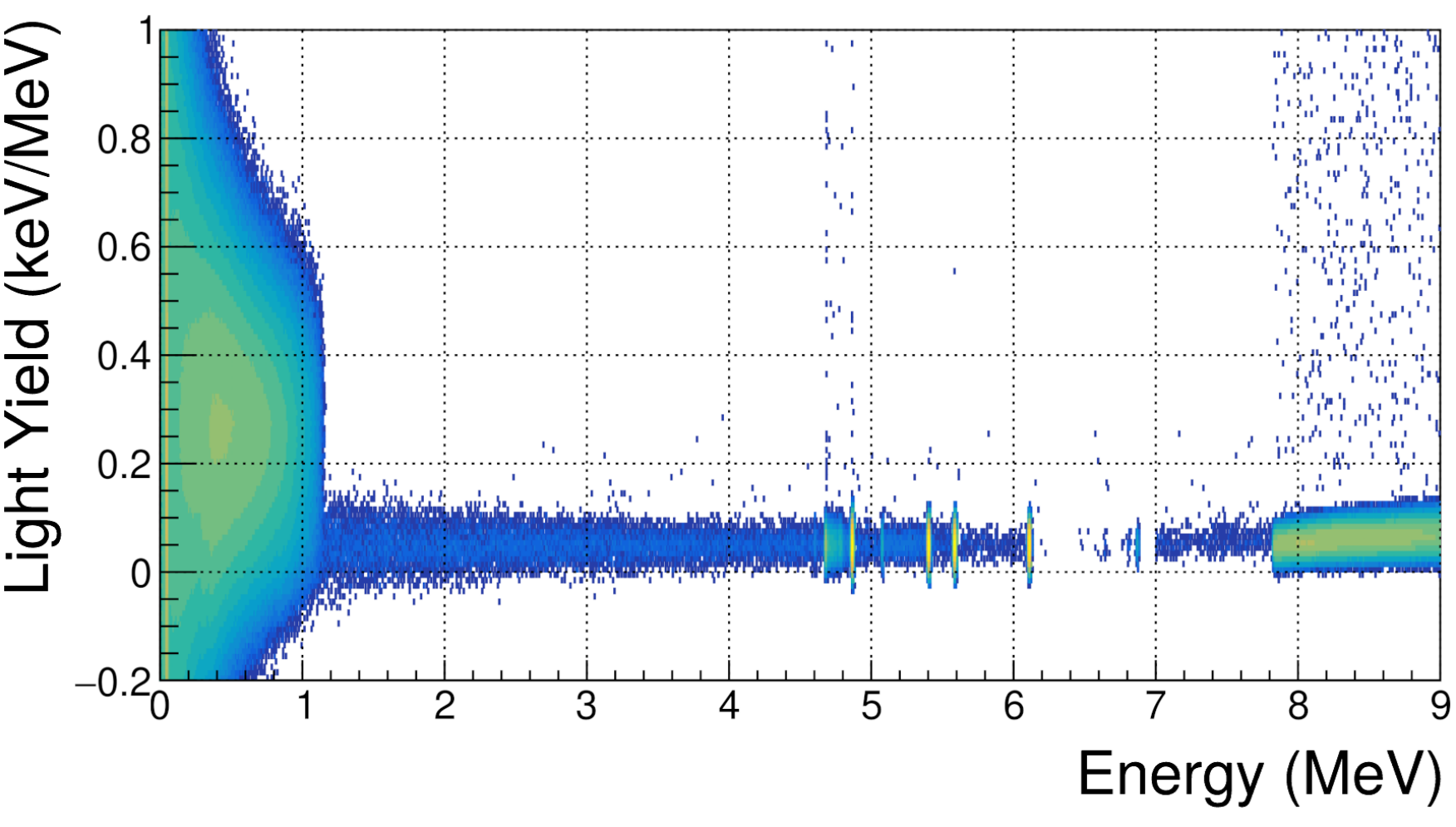}
  \includegraphics[width=0.49\linewidth]{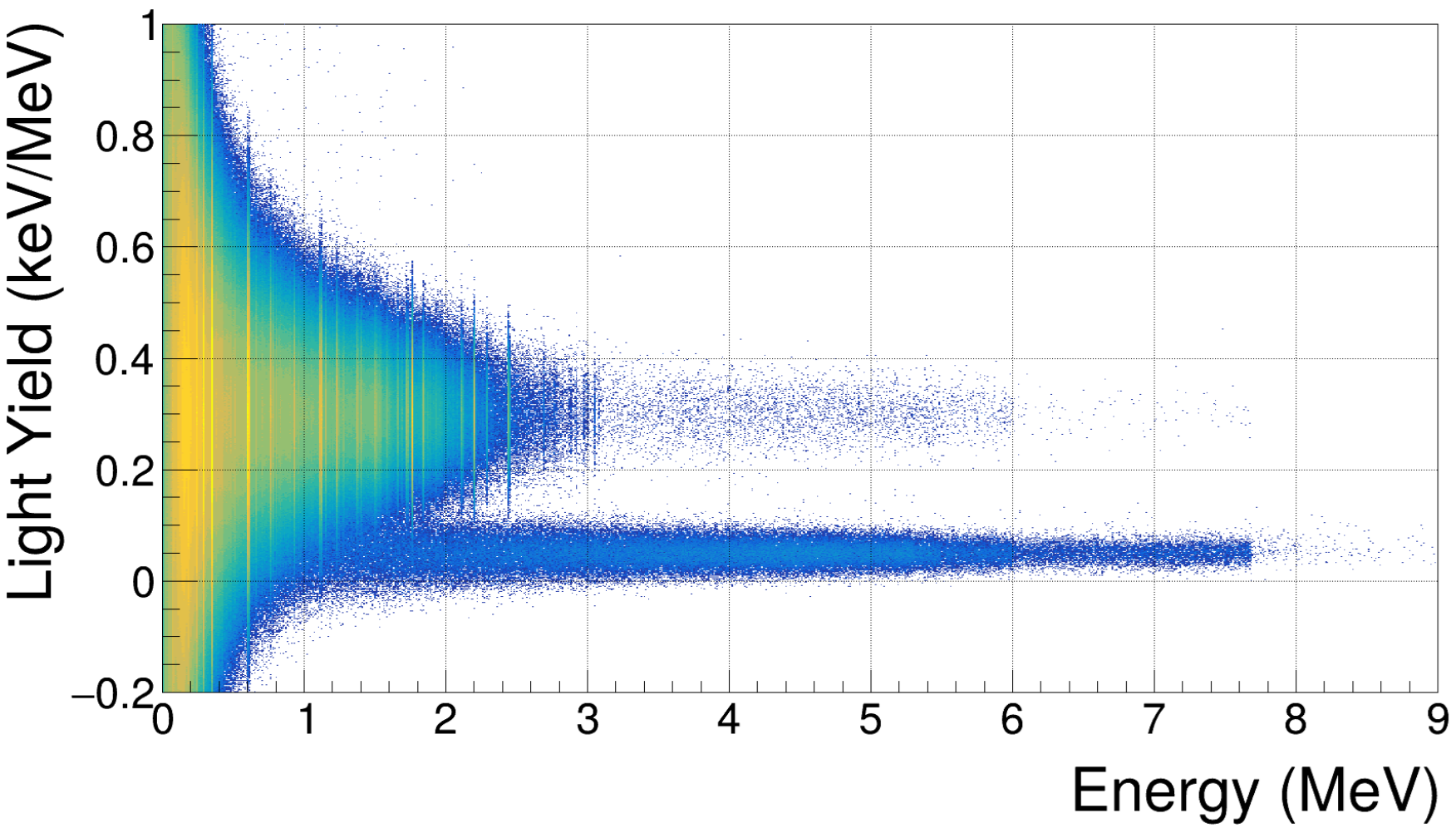}
  \includegraphics[width=0.49\linewidth]{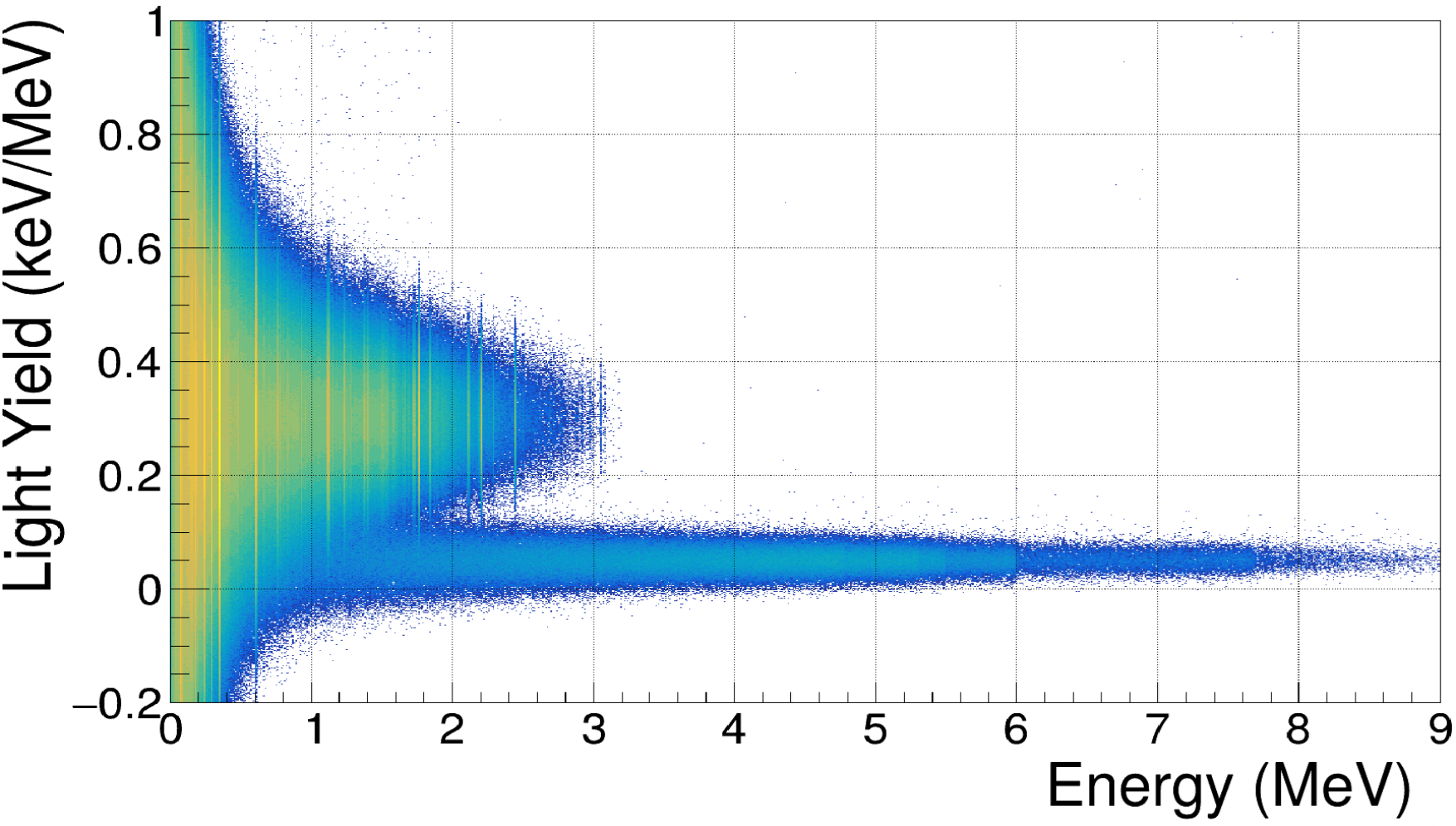}
  \includegraphics[width=0.49\linewidth]{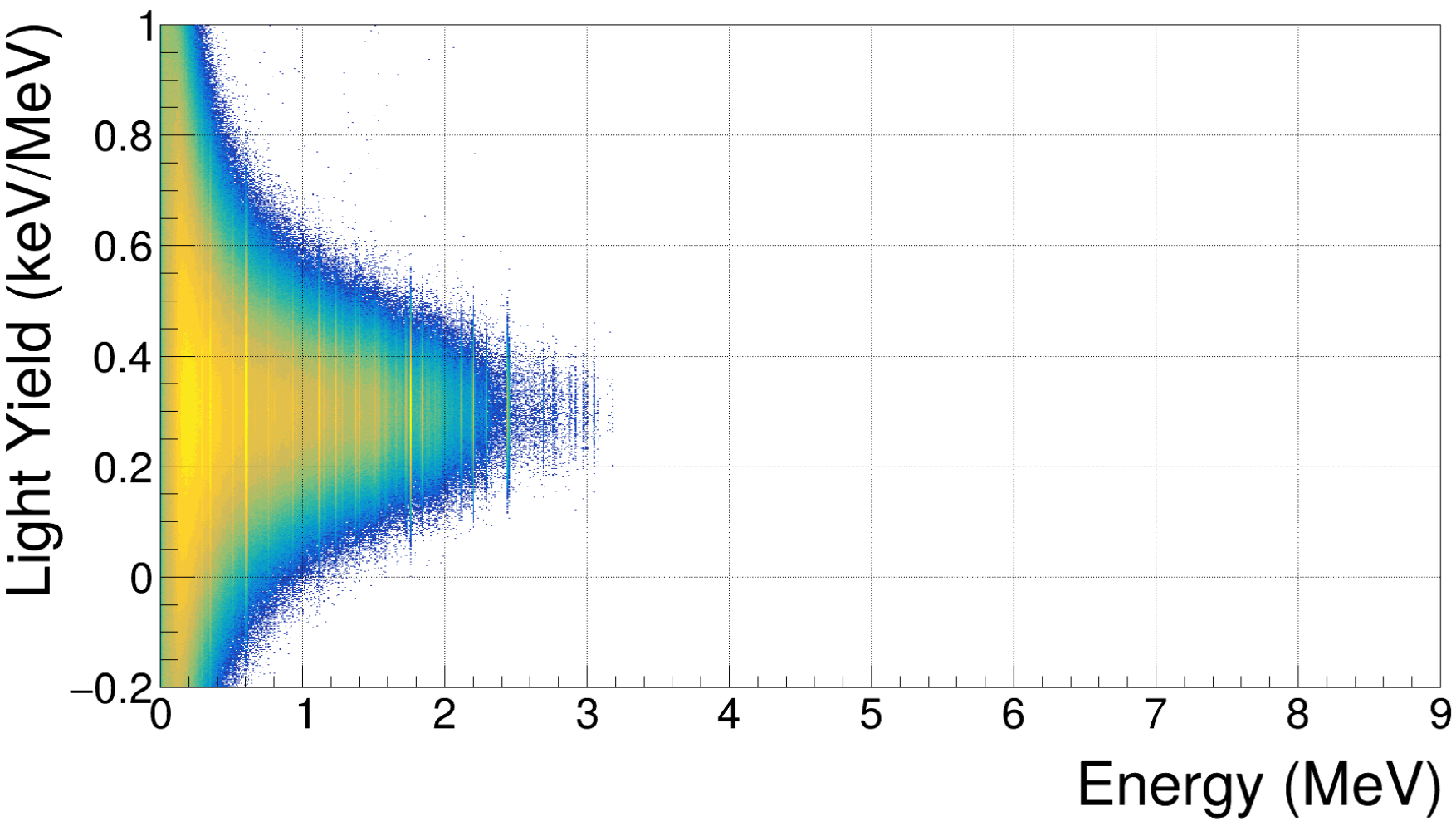}
  \caption{Scatter plots of the LHR parameter versus the energy deposited in LMOs obtained in simulations of the decay sub-chain of the $^{226}$Ra isotope (up to $^{210}$Pb) homogeneously distributed in four different volumes: the LMO crystals (top left), the 10~mK screen (top right), the copper frames (bottom left) and the external lead shielding (bottom right).  
  Present particle identification capability can be further improved using the NTL amplification of LD signals, as foreseen in CROSS.}
  \label{fig:Ly(H)Sim}
\end{figure*}

After summing the energy of scintillation light to the energy deposited in the LD and considering the energy resolution of these detectors, it is possible to build the same scatter plots as for the experimental data. Figure~\ref{fig:Ly(H)Sim} shows the LHR as a function of the energy deposited in the crystals for the decay chain of the $^{226}$Ra isotope (up to the $^{210}$Pb) in four different volumes: the crystals themselves, the 10~mK Cu screen, the copper frames and the external lead shielding. In simulation of the background in the crystals, it is possible to observe the characteristic monoenergetic peaks in the $\alpha$ band, which are produced by those decays that deposit all the energy in the LMO, and the low energy $\alpha$ events are produced when these particles deposit part of the energy in the crystal and then escape from the thermal detector. It is worth noticing that the events in the $\alpha$ band with energies greater than 7.8~MeV are produced by coincidence within the 1~ms integration window of the Bi-Po chain. In these cases, both the energies of the $\alpha$ and $\beta$ particles are summed in a single event. Both the frames and the screens face the LMOs; therefore, the $\alpha$ particles emitted in the decays of this chain can reach the detectors with an energy equal to or lower than the initial energy of the $\alpha$ particle. Finally, events in the $\beta$/$\gamma$ band with energies between 3~and 5~MeV present in the 10~mK screen simulation are secondary particles emitted in the volumes surrounding the detectors after an $\alpha$ interaction. Thus, radioactive contamination of the 10 mK screen by radium and thorium is the most harmful background for CROSS, as it falls within the ROI.

%---------------------------------------------------------
\subsection{Background sources}
\label{sec:Sources}

We considered different background sources which might contribute to the $^{100}$Mo ROI of the CROSS experiment, such as:
\begin{itemize}
    \item naturally occurring radioisotopes of U/Th decay chains, present in the construction elements of the detector and the facility;
    \item $2\nu2\beta$ decay of $^{100}$Mo in LMO crystals;
    \item random coincidences, mainly two $^{100}$Mo $2\nu2\beta$ decay events, given a comparatively high $^{100}$Mo $2\nu2\beta$ rate (3~mHz) in the CROSS crystals and a slow response of the thermal detectors, with a characteristic pulse rise time of $O$(1--10 ms);
    \item cosmic muons at the experimental location.
\end{itemize}
 
Thermal neutron capture in materials surrounding the detector components and in the detectors themselves can lead to $\gamma$ emissions with energies extending up to several MeV, potentially overlapping with the 0$\nu2\beta$ ROI of $^{100}$Mo. However, the neutron-induced background is expected to be strongly suppressed in the CROSS experiment due to the underground location of the setup, the presence of massive passive shielding, and the relatively low environmental neutron flux in the laboratory reported in the literature. Previous measurements performed within the CROSS setup \cite{CROSSdeplLMO:2023} indicate a thermal neutron detection rate of approximately 2 counts/day in LMO crystals, confirming the low neutron flux at the experimental site. Given these considerations, neutron-induced backgrounds are not expected to represent a dominant contribution to the total background index. Nevertheless, their potential impact will be further investigated in future dedicated simulations and measurements.

%---------------------------------------------------------
\subsubsection{Radioactive decay chains}

Among the natural decay chains, only several radionuclides of $^{232}$Th, $^{235}$U, and $^{238}$U can contribute to the background in the ROI. Since contamination of isotopes from the $^{235}$U decay chain in the materials used in the experiment is negligible compared to the two other decay chains, this family has not been simulated. The other two decay chains are used to be out of equilibrium as a result of component refinement. Therefore, they have been simulated from $^{226}$Ra to $^{210}$Pb for the uranium decay chain and from $^{228}$Th to $^{208}$Pb for the thorium one using initial decay kinematics provided by the G4RadioactiveDecay class of Geant4.

The $^{226}$Ra and $^{228}$Th decay sub-chains have been simulated in all volumes included in the simulated geometry, and the primary particles were homogeneously distributed in the bulk of the volumes. It is worth noticing that this simulation does not consider any special contribution to the radioactive components in the surface of the materials. Although the geometry of the electronic pins was not simulated, their contribution to the background has been considered since they are one of the most contaminated elements of the set-up. They have been simulated as point-like sources in the positions of the electronic pins on top of the internal lead shielding.

The activities of $^{226}$Ra and $^{228}$Th in all simulated volumes are listed in Table~\ref{tab:Activities}. Radioactive contaminations of copper cryostat components, electronic pins, lead shielding, fiberglass bars, and Kapton pads were measured using HPGe $\gamma$ spectroscopy. The activities of $^{226}$Ra and $^{228}$Th in the bulk of the crystals and the copper structure of the CROSS detector array were obtained from the background model of the CUPID-Mo experiment \cite{Augier:2023model}, given that the same materials with similar characteristics and from the same producers are used. 
Since only the upper limits at 90\% of confidence level (C.L.) are found for the $^{226}$Ra and $^{228}$Th activities in the lead shielding, they have been simulated considering these upper limits as input values.

\begin{table}
    \centering
    \caption{Activities of $^{226}$Ra and $^{228}$Th in the simulated volumes of the CROSS experiment. Radioactive contaminations of Cu screens of the cryostat, electronic pins, lead shielding, fiberglass bars, and Kapton pads were screened using HPGe $\gamma$ spectroscopy. Radiopurity of the enriched crystals and the copper frames is assumed to be the same as reported by the CUPID-Mo experiment \cite{Augier:2023model}.}
    \label{tab:Activities}
    \begin{tabular}{c|c|c}
    \hline
        Material & $^{226}$Ra ($\mu$Bq/kg) & $^{228}$Th ($\mu$Bq/kg) \\ \hline
        Copper screens & 0.6(1) $\times$ 10$^3$ & 0.3(1) $\times$ 10$^3$ \\
        Electronic pins & 1.33(4) $\times$ 10$^6$ & 2.39(3) $\times$ 10$^6$ \\
        Crystals (bulk) & 0.39(6) & 0.57(7) \\
        Copper frames & 25(15) & 33(16) \\
        Lead shielding & $<$ 120 & $<$ 460\\
        Fiberglass bars & 3.4(4) $\times$ 10$^3$ & 1.41(5) $\times$ 10$^3$ \\
        Kapton pads & 17(7) $\times$ 10$^3$  & 67(31) $\times$ 10$^3$ \\
        \hline
    \end{tabular}
\end{table}

%---------------------------------------------------------
\subsubsection{$2\nu2\beta$ decay of $^{100}$Mo}

One of the most important contributions to the background below the $0\nu2\beta$ ROI is expected to be the $2\nu2\beta$ decays of $^{100}$Mo \cite{Chernyak:2012}. To simulate the energy deposited in each detector after these decays, a new method has been developed using Geant4 and implemented in the following steps:
\begin{enumerate}
    \item The energy and momentum vector of the two electrons are obtained using the Decay0 event generator \cite{Ponkratenko:2000}. Only the decay of the $^{100}$Mo isotope to the ground state of $^{100}$Ru has been simulated (the rate of the $2\nu2\beta$ transition to the excited state is 2 orders of magnitude lower \cite{Pritychenko:2025,Augier:2023excited}).
    \item A random position inside the LMO crystals is selected.
    \item Both electrons are simulated using Geant4 as primary particles with the momentum obtained in step 1 and the position obtained in step 2.
    \item The energy deposited in each detector for both simulated electrons is summed and all the information is combined to generate a single event.
\end{enumerate}

Once the $2\nu2\beta$ event is generated with the information provided by Geant4, it is possible to apply the same event selections across the different detectors as in the simulations of the radioactive decay chains. 

In the same way as for the $2\nu2\beta$ decays, the $0\nu2\beta$ decays have also been simulated, since it is the signal of our experiment. The lifetime considered to calculate the activity and normalize the energy spectrum was the current most stringent limit set on this decay for $^{100}$Mo (2.9 $\times$ 10$^{24}$~yr \cite{Agrawal:2025amoreI}). 
This allowed us to compare the signal with the background and to obtain the signal event selection efficiency and 
to calculate a limit on the $^{100}$Mo $0\nu2\beta$ process. 
We found that among all the simulated $0\nu2\beta$ decays, 87\% of them are single-crystal hits and 78\% of them are single-crystal hit events with the full energy deposit in the LMO crystal.

%---------------------------------------------------------
\subsubsection{Pile-up of $2\nu2\beta$ from $^{100}$Mo}

It can happen that two non-related interactions occur in the same detector so close in time that they cannot be distinguished as separate signals. In such cases, the pulse-shape discrimination (PSD) parameter cannot identify the event pile-up, and the two events are considered in the analysis as a single one with an energy that is the sum of the energy deposited in both events. The rate of this process is the product of the rates of both piled-up events and the maximum time up to which they can be resolved. Due to the slow time response of LDs (rise time of $\sim$1~ms), the contribution of these pile-up events is non-negligible~\cite{CROSSpileup:2023}. This process is more important for the events with the highest rate in the experiment, i.e. the $^{100}$Mo $2\nu2\beta$ decays. Moreover, since the maximum value of the energy release in the $2\nu2\beta$ decay of $^{100}$Mo is 3034~keV, the energy distribution of the pile-up events spreads to 6~MeV, contributing to the ROI.

This background contribution has been simulated in a similar process as that followed for the $2\nu2\beta$ decay of $^{100}$Mo, except that now 4 electrons have been integrated in the same event. Although the performance of the LDs as reported in \cite{CROSS_Run9:2025} 
shows that pile-up events can be rejected when they are separated by $\sim$0.5~ms, in this simulation a wider time window was considered to account for other slower detectors. We assumed that pile-ups cannot be identified when two $^{100}$Mo nuclei decay in a time window of 1~ms, the scale of the LD rise time. The resulting energy spectrum of the pile-up events has been normalized considering that their rate is the product of this time window times the square of the $^{100}$Mo $2\nu2\beta$ decay rate.

%---------------------------------------------------------
\subsubsection{Muons}

The background contribution of muons and secondary particles produced after their interaction with the volumes of the experiment is expected to be dominant due to the relatively high muon rate in the LSC (around 20~$\mu$/m$^2$/h)~\cite{Trzaska:2019}. 
Therefore, in addition to the muon veto, constructed around the CROSS facility, a dedicated investigation was realized to optimize event selections across the different detectors to minimize the contribution of muon-induced events. This work is described in detail in \cite{CROSS_MuonVeto:2026}, while a short overview is given below.

The initial position of the muons is a flat surface on top of the experimental set-up. This surface covers a diameter enough to simulate azimuthal angles up to 75$^o$. The initial momentum and position of the muons are randomly selected from the 2D angular distribution measured in the laboratory in \cite{Trzaska:2019}. The energy of each simulated muon is randomly selected from a PDF given by 
\begin{equation}
    \frac{dN_{\mu}}{dE_{\mu,0}~d\Omega} \approx \frac{0.14~E_{\mu,0}^{-2.7}}{cm^2~s~sr~GeV} \times \left[\frac{1}{1+\frac{1.1E_{\mu,0}~\cos\theta}{115~GeV}} + \frac{0.054}{1+\frac{1.1E_{\mu,0}~\cos\theta}{850~GeV}}\right].
    \label{eq:muonEnergy}
\end{equation}
This equation is an approximation of the muon energy spectrum above ground and is reliable for $E_{\mu,0} >$~100~GeV. Since all the muons reaching the laboratory should have much larger energies aboveground, this approach is very suitable for this simulation. Once the energy of a muon was obtained above ground, $E_{\mu,0}$, the energy of that muon in the laboratory was calculated as 
\begin{equation}
    E_{\mu} = (E_{\mu,0} + \epsilon)\cdot e^{-bX}-\epsilon ,
    \label{eq:muonEnergyLab}
\end{equation}
where $X$ is the slant depth and $b^{-1}\approx$~2500~m~w.e. and $\epsilon\approx$~2.5~TeV are the rock overburden parameters. Only muons satisfying 
\begin{equation}
    E_{\mu,0} > \epsilon\cdot (e^{bX}-1),
    \label{eq:muonEnergyLab2}
\end{equation}
were considered in the simulation to ensure a positive energy at the laboratory depth. Muons below this threshold are absorbed in the overburden before reaching the laboratory.

The energy spectrum in the detectors is normalized considering the muon flux measured in the laboratory \cite{Trzaska:2019}. 
Events have been grouped into two categories: those in which the muons interacted with the crystal and those in which all the energy was deposited by secondary particles produced after the muon interaction with the materials surrounding the detectors. The contribution of secondary particles to the background in the ROI of the experiment was found to be $\sim$80\% of muon-induced events. These secondary particles are mainly $\gamma$'s and are mostly produced in the lead shielding.

All events in thermal detectors with an energy deposit in the muon veto larger than the threshold set in the experiment (1.5~MeV) in a time window of 2~ms have been rejected from the analysis, in the same way as will be done in the experimental data.

\begin{figure*}[!ht]
    \centering
    \includegraphics[width=0.9\linewidth]{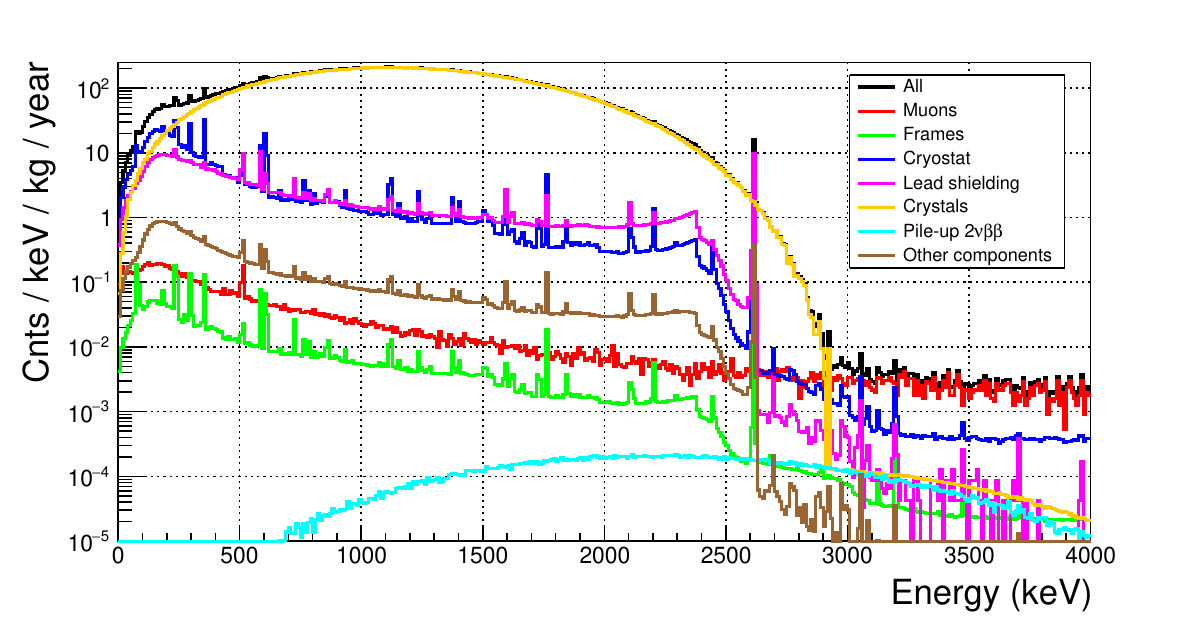}
    \caption{Background energy spectra produced by each simulated volume of the CROSS experiment. The spectrum labeled ``Crystals'' includes the contributions of the $^{226}$Ra and $^{228}$Th decay chains, and the $2\nu2\beta$ decay of $^{100}$Mo. The spectrum labeled ``Other components'' includes the contribution from the electronic pins, the fiberglass bars and the Kapton pads.    }
    \label{fig:BackgroundSpectra}
\end{figure*}

\begin{figure*}[!ht]
    \centering
    \includegraphics[width=1.0\linewidth]{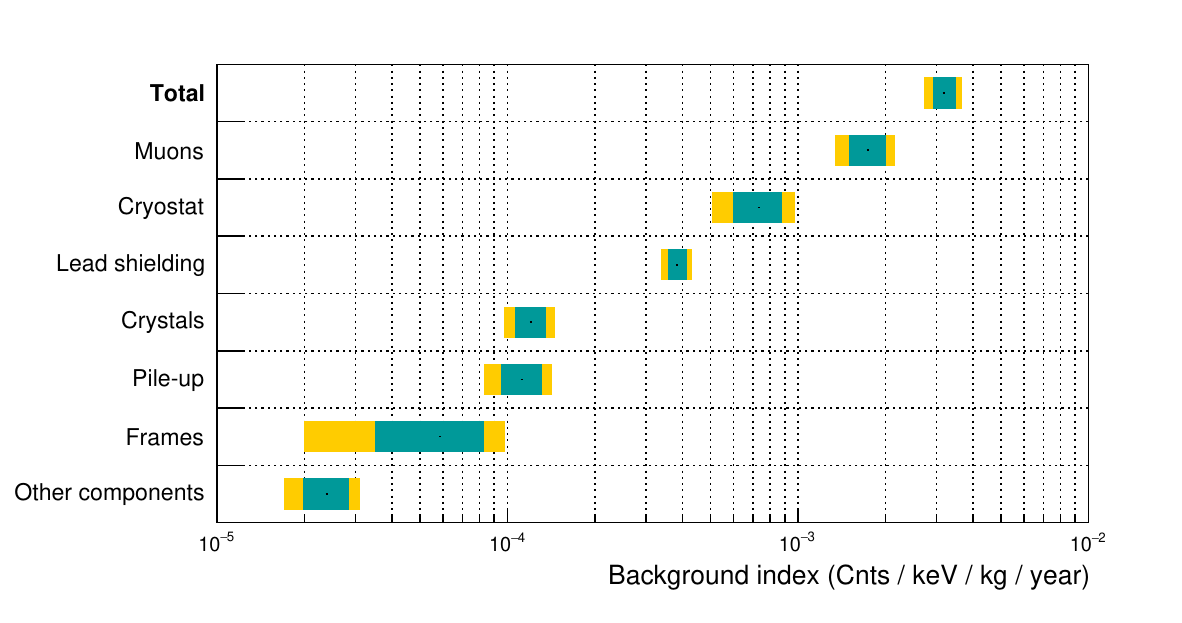}
    \caption{Background indices in the ROI for each component of the CROSS experiment considered in the simulation. The label "Other components" includes the background contribution from the electronic pins, the fiberglass bars and the Kapton pads.}
    \label{fig:BackgroundIndex}
\end{figure*}

%---------------------------------------------------------
\subsection{Event selection}
\label{sec:EventSelection}

In order to use an event in the analysis for $0\nu2\beta$ of $^{100}$Mo, it must meet a few conditions. These requirements are built to minimize the presence of background events in our analysis:
\begin{enumerate}

    \item Since the $2\beta$ decay is localized inside just one crystal, the event must be a single-crystal hit.
    
    \item The event must not be in coincidence with any veto module. As written above, the time window used for this coincidence is 2~ms.
    
    \item The LHR parameter of that event must fit inside the $\gamma$($\beta$) band.
    
    \item There must not be other energy deposits within the time window used for the analysis of a given pulse (usually 1~ms). This event rejection is usually performed in experimental data with a PSD parameter that compares the pulse of each event with respect to the mean pulse. If there is more than one energy deposit in the time window used in the analysis, the pulse shape of that event becomes different from that of the reference event.
    
\end{enumerate}

%---------------------------------------------------------
\subsection{Background index in the region of interest}
\label{sec:BI}

After the simulation of all the background contributions, the event selections presented in section~\ref{sec:EventSelection} have been applied. To build the energy spectra, the 36~LMOs of the CROSS array have been considered as a single detector whose mass is the sum of the masses of all the LMOs. Figure~\ref{fig:BackgroundSpectra} shows the background energy spectra produced by each simulated volume. The crystal bulk activity related spectrum includes both the contributions of the $^{226}$Ra and $^{228}$Th decay sub-chains and the $2\nu2\beta$ from $^{100}$Mo. The spectrum named “Other components” includes electronic pins, fiberglass bars, and Kapton pads.

The contribution to BI in ROI for each component has been calculated by integrating the events in each spectrum in a 100-keV-wide energy range centered at the Q$_{2\beta}$ value of $^{100}$Mo. The results for each contribution are shown in Fig.~\ref{fig:BackgroundIndex}. The uncertainties given for each contribution were calculated taking into account the uncertainties of the activities presented in Table~\ref{tab:Activities} and the statistical uncertainties of the number of simulated events. It should be noted that the most important contribution to BI is produced by muon-induced events. Then, with the background model explained in this section, the total BI expected for the experiment is 3.2(5)~$\times$~10$^{-3}$~cnts/keV/kg/yr.

%%=======================================================
\section{CROSS sensitivity to $^{100}$Mo $0\nu2\beta$ decay}
\label{sec:sensitivity}

Using the MC prediction of the background expected in the CROSS experiment at the $^{100}$Mo Q$_{2\beta}$ $\pm$ 50~keV interval, 3.2(5) $\times$ 10$^{-3}$ cnts/keV/kg/yr, we can estimate the experimental sensitivity to $0\nu2\beta$ decay in $^{100}$Mo. 
We assume an ROI width of 17.1 keV, as achieved in CUPID-Mo with similar performance LMO detectors \cite{Augier:2023}, and we take the total mass of LMO crystals (8.93 kg) in the CROSS array to estimate the number of background counts in the ROI. 
Taking into account the $^{100}$Mo enrichment of molybdenum used for the LMO growth, the number of $^{100}$Mo nuclei embedded in the CROSS detectors is 2.95 $\times$ 10$^{25}$. 
The $0\nu2\beta$ signal containment efficiency in the ROI, computed in our MC simulations, is 78\%, while the efficiency of selection cuts is assumed to be 90\% (similarly to CUPID \cite{CUPID_baseline:2025}), resulting in a total efficiency of 70.2\%. 
The elapsed time of the CROSS experiment is 1 year (2026), with a possible extension to up to 2 years, thus, we consider a 3-yr-long period in our analysis. 
The duty cycle of the experimental setup is assumed to be 90\%, which is usually achieved in the long cryogenic runs of the CROSS facility \cite{CROSSdetectorStructure:2024}. 
Given a comparatively high trigger rate of the muon veto system, $O$(100~Hz), we expect a quite large portion of the dead time induced by anti-coincidences with the veto: 18\% (estimated in muon simulations and validated with low-temperature measurements \cite{CROSS_MuonVeto:2026}). 
Having background counts in the ROI for different elapsed time periods, we use  
the Feldman-Cousins procedure \cite{Feldman:1998} to compute the number of events of the $0\nu2\beta$ effect searched for excluded at 90\% C.L. 
Finally, combining together the number of $^{100}$Mo nuclei, the total efficiency, the live time, and the number of $0\nu2\beta$ events excluded at 90\% C.L., we estimate the $^{100}$Mo half-life limit relative to the $0\nu2\beta$ decay.

\begin{figure*}[!ht]
    \centering
    \includegraphics[width=1.0\linewidth]{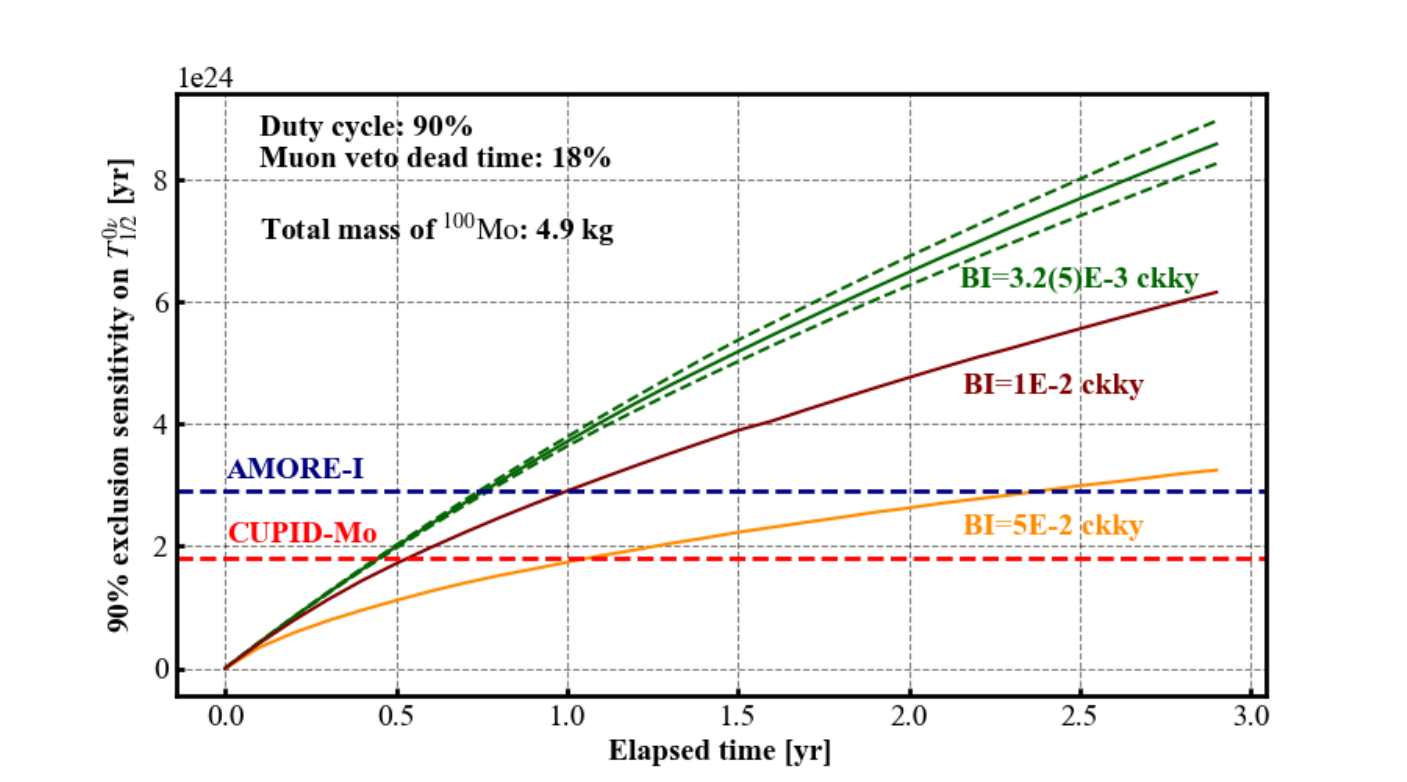}
    \caption{Sensitivity of the CROSS experiment to $^{100}$Mo $0\nu2\beta$ decay with the elapsed time of data taking. A duty cycle of 90\% and a muon veto induced dead time of 18\% are assumed according to results of previous cryogenic runs of the CROSS setup and dedicated simulations. The background index in ROI, 3.2(5) $\times$ 10$^{-3}$ cnts/keV/kg/yr (ckky), corresponds to the background projections of the present work (Sect. \ref{sec:background}). Also, we plot the CROSS sensitivity for a conservative BI value, 1 $\times$ 10$^{-2}$ cnts/keV/kg/yr, considered in the CROSS proposal \cite{Bandac:2020} and the extreme worst-case scenario assuming BI = 5 $\times$ 10$^{-2}$ cnts/keV/kg/yr. The half-life limits reported by the CUPID-Mo (1.5 kg$\times$yr of $^{100}$Mo exposure) \cite{Augier:2022} and AMoRE-I (4.9 kg$\times$yr) \cite{Agrawal:2025amoreI} experiments are given for comparison.}
    \label{fig:Sensitivity}
\end{figure*}

The projected exclusion sensitivity with the elapsed time of the CROSS experiment is shown in Fig. \ref{fig:Sensitivity}. This projection demonstrates the prospects for the CROSS experiment to set the most stringent limit on $^{100}$Mo $0\nu2\beta$ decay in less than 1 year of data collection, if the BI of 3.2(5) $\times$ 10$^{-3}$ cnts/keV/kg/yr will be achieved (a similar level has been already demonstrated by CUPID-0 \cite{Azzolini:2019nmi} and CUPID-Mo \cite{Augier:2023model}). However, taking into account that radioactivity of materials is unpredictable to a certain extent (and also the energy resolution of the detectors), we present the sensitivity also in much more conservative options: 
BI = 1 $\times$ 10$^{-2}$ cnts/keV/kg/yr, as considered in the CROSS proposal \cite{Bandac:2020}; 
BI = 5 $\times$ 10$^{-2}$ cnts/keV/kg/yr, a factor 2 worse than the background in ROI achieved in the AMoRE-I experiment \cite{Agrawal:2025amoreI}. One can see in Fig. \ref{fig:Sensitivity} that the conservative case of BI = 1 $\times$ 10$^{-2}$ cnts/keV/kg/yr would still allow to reach the world-best sensitivity to $0\nu2\beta$ decay in $^{100}$Mo over about 1.5 yr of data taking, while for the worst-case scenario corresponding to BI = 5 $\times$ 10$^{-2}$ cnts/keV/kg/yr, the CROSS experiment would need to run about 2 years to be reasonably competitive.

%%=======================================================
\section{Conclusions}
\label{sec:conclusions}

We present a detailed description of the CROSS detector array made of 42 thermal detector modules with dual heat-light readout, 32 of which are based on $^{100}$Mo-enriched lithium molibdate (LMO) crystal scintillators (0.28 kg each) to search for $0\nu2\beta$ decay in $^{100}$Mo. The light detection is provided by thin Ge or Si thermal detectors with voltage-driven signal amplification based on Neganov-Trofimov-Luke effect. The detector array has been assembled in a clean room of IJCLab (Orsay, France), using a specially developed mechanical structure, which provides a very low ratio between Cu (the main construction element) and LMO crystals, only 6\%. The array was transported to Spain and installed in the Canfranc underground laboratory, in a dedicated low-background cryogenic facility that has been operational since 2019. The CROSS detector array was cooled to $\sim$20 mK in October 2025 and commissioned between November 2025 and March 2026; regular physics data taking has been ongoing since April 2026.

We also describe a Geant4-based model of the CROSS experiment and dedicated Monte Carlo simulations of backgrounds induced by environmental radioactivity and components of the cryogenic facility and the detector array. We found that all sources but cosmic muons contribute below $\sim$$10^{-4}$--$10^{-3}$ cnts/keV/kg/yr, and the total background index in a 100-keV-wide interval centered at the $^{100}$Mo Q$_{2\beta}$ is predicted to be 3.2(5) $\times$ 10$^{-3}$ cnts/keV/kg/yr. 

Given the MC-based background projection in the CROSS experiment, the number of $^{100}$Mo nuclei embedded in detectors ($2.95 \times 10^{25}$), assuming the width of the region of interest 17 keV, the total efficiency 70\%, and live time 90\% with the muon veto system revealing a dead time of 18\%, the CROSS experiment has the potential to reach the world-leading sensitivity to $^{100}$Mo $0\nu2\beta$ decay in less than a year of data collection. After 2 years the exclusion sensitivity to $^{100}$Mo $0\nu2\beta$ decay is expected to be about 6 $\times$ $10^{24}$~yr at 90\% C.L. 
Assuming conservatively a factor 3 (10) worse background index due to unpredictable materials' radiopurity and/or detector performance, the running of CROSS experiment over 2 years would still allow to reach the best (competitive) sensitivity to $0\nu2\beta$ decay in $^{100}$Mo.

%%=======================================================

\section{Acknowledgments}
This work is supported by the European Commission (Project CROSS, Grant No. ERC-2016-ADG, ID 742345) and by the Agence Nationale de la Recherche (Project CUPID-1; ANR-21-CE31-0014, ANR France). We also acknowledge the support of P2IO LabEx (ANR-10-LABX0038) in the framework ”Investissements d’Avenir” (ANR-11-IDEX-0003-01 – Project ”BSM-nu”) managed by ANR, France. 
Russian and Ukrainian scientists have given and give crucial contributions to CROSS. For this reason, the CROSS collaboration is particularly sensitive to the current situation in Ukraine. The position of the collaboration leadership on this matter, approved by majority, is expressed in \href{https://a2c.ijclab.in2p3.fr/en/a2c-home-en/assd-home-en/assd-cross/}{https://a2c.ijclab.in2p3.fr/en/a2c-home-en/assd-home-en/assd-cross/}.

%%=======================================================

\bibliographystyle{spphys}       % APS-like
%\bibliography{Bibliography}

\end{document}